\documentclass[acmlarge]{acmart}
\makeatletter
\newcommand{\confshort}{\acmConference@shortname}
\newcommand{\conffull}{\acmConference@name}
\newcommand{\confdate}{\acmConference@date}
\newcommand{\confloc}{\acmConference@venue}
\AtBeginDocument{
  \fancypagestyle{firstpagestyle}{
    \fancyhead{}%
    \fancyfoot[C]{}%
  }
  \fancyhf{}
  \fancyhead[LO]{\@headfootfont\shorttitle}%
  \fancyhead[RE]{\@headfootfont\@shortauthors}%
  \fancyhead[LE]{\@headfootfont\footnotesize \confshort, \confdate, \confloc}%
  \fancyhead[RO]{\@headfootfont\footnotesize \confshort, \confdate, \confloc}%
  \fancyfoot[C]{}%
}
\makeatother
\acmBooktitle{\conffull\@ (\confshort), \confdate, \confloc}

\AtBeginDocument{%
  \providecommand\BibTeX{{%
    Bib\TeX}}}

\setcopyright{acmlicensed}
\copyrightyear{2026}
\acmYear{2026}
\setcopyright{cc}
\setcctype{by-nc-nd}
\acmConference[FAccT '26]{The 2026 ACM Conference on Fairness, Accountability, and Transparency}{June 25--28, 2026}{Montreal, QC, Canada}
\acmBooktitle{The 2026 ACM Conference on Fairness, Accountability, and Transparency (FAccT '26), June 25--28, 2026, Montreal, QC, Canada}
\acmDOI{10.1145/3805689.3806763}
\acmISBN{979-8-4007-2596-8/2026/06}

\usepackage{graphicx}
\usepackage{subcaption}
\usepackage{tabularx}
\usepackage{amsmath}
\usepackage{xcolor}
\usepackage[normalem]{ulem}
\usepackage{multirow}
\usepackage{longtable}
\usepackage{float}
\usepackage{booktabs}

\usepackage{stfloats}
\usepackage{enumitem}

\usepackage{tcolorbox}
\tcbuselibrary{skins}
\definecolor{remarkpurple}{RGB}{138,42,123}
\definecolor{lightpurple}{RGB}{247,235,245}
\newtcolorbox{purplebox}[1][]{
    colback=lightpurple,
    colframe=remarkpurple,
    rounded corners,
    boxrule=1.5pt,
    left=1pt,
    right=1pt,
    top=1pt,
    bottom=1pt,
    #1
}

\definecolor{remarkpink}{RGB}{209,45,124}
\definecolor{lightpink}{RGB}{254,240,248}
\newtcolorbox{pinkbox}[1][]{
    colback=lightpink,
    colframe=remarkpink,
    rounded corners,
    boxrule=1.5pt,
    left=1pt,
    right=1pt,
    top=1pt,
    bottom=1pt,
    #1
}

\definecolor{remarkred}{RGB}{206,45,37}
\definecolor{lightred}{RGB}{254,240,240}
\newtcolorbox{redbox}[1][]{
    colback=lightred,
    colframe=remarkred,
    rounded corners,
    boxrule=1.5pt,
    left=1pt,
    right=1pt,
    top=1pt,
    bottom=1pt,
    #1
}

\definecolor{remarkorange}{RGB}{220,110,44}
\definecolor{lightorange}{RGB}{255,245,237}
\newtcolorbox{orangebox}[1][]{
    colback=lightorange,
    colframe=remarkorange,
    rounded corners,
    boxrule=1.5pt,
    left=1pt,
    right=1pt,
    top=1pt,
    bottom=1pt,
    #1
}

\definecolor{remarkyellow}{RGB}{241,188,65}
\definecolor{lightyellow}{RGB}{255,252,240}
\newtcolorbox{yellowbox}[1][]{
    colback=lightyellow,
    colframe=remarkyellow,
    rounded corners,
    boxrule=1.5pt,
    left=1pt,
    right=1pt,
    top=1pt,
    bottom=1pt,
    #1
}

\definecolor{keywordcolor}{RGB}{0, 102, 204}
\definecolor{operatorcolor}{RGB}{204, 0, 0}
\definecolor{termcolor}{RGB}{0, 128, 0}
\definecolor{extracolor}{RGB}{210, 39, 119}

\newtoggle{comments}  %
\settoggle{comments}{true}
\iftoggle{comments}{
    \newcommand{\notejs}[1]
    	{{\color{brown}[{\bf Jat:} #1]}}
	\newcommand{\notean}[1]
    	{{\color{violet}[{\bf Anna:} #1]}}
    \newcommand{\notehs}[1]
    	{{\color{magenta}[{\bf Holli:} #1]}}
    \newcommand{\notebz}[1]
        {{\color{blue}[{\bf Bilal:} #1]}}
        \newcommand{\todo}[1]
    	{{\color{red}[{\bf Todo:} #1]}}

}{
    \newcommand{\notejs}[1]{} 
	\newcommand{\notean}[1]{}
    \newcommand{\notehs}[1]{}	
    \newcommand{\notebz}[1]{}
    \newcommand{\todo}[1]{}

}
\AtBeginDocument{%
   \providecommand\BibTeX{{%
     \normalfont B\kern-0.5em{\scshape i\kern-0.25em b}\kern-0.8em\TeX}}}

\begin{document}

\title{Prompt Governance? On Governing Technologies Governed by Natural Language}

\author{Anna Neumann}
\affiliation{
    \institution{Research Centre Trust, UA Ruhr, University of Duisburg-Essen}
    \city{Duisburg}
    \country{Germany}
}
\email{anna.neumann1@uni-due.de}

\author{Holli Sargeant}
\affiliation{
        \institution{University of Cambridge}
        \city{Cambridge}
        \country{United Kingdom}
    }
\email{hs775@cam.ac.uk}

 \author{Jatinder Singh}
    \affiliation{
        \institution{Research Centre Trust, UA Ruhr, University of Duisburg-Essen}
        \city{Duisburg}
        \country{Germany}
    }
    \affiliation{
        \institution{University of Cambridge}
        \city{Cambridge}
        \country{United Kingdom}
    }
    \email{jat@compacctsys.net}

\renewcommand{\shortauthors}{Neumann et al.}

\begin{abstract}
Generative artificial intelligence (GenAI) is increasingly operated by natural language instructions (prompts). Across the pipeline, stakeholders designate various forms of natural language instructions, e.g., end-user guidelines, developer specifications, or system prompts, as \textit{prompt governance} instruments. These textual artifacts are intended to shape model behaviour by specifying constraints, priorities, and compliance rules.
Policymakers and regulators have thus begun to treat system-level instructions (system prompts) as accessible prompt-based GenAI intervention points, assuming they function (directly or indirectly) as behavioural controls. 
Yet whether these instructions operate reliably, consistently, and predictably enough across contexts and adversarial settings to support such governance frameworks remains under-explored.
Towards this, we systematically evaluate (i) how researchers discuss and treat system-level instructions in the existing literature, focusing on large language models (LLMs) as they isolate language effects, (ii) how policymakers position system-level instructions as governance objects, incorporating analysis of two recent policy frameworks (Executive Order 14319 `Preventing Woke AI in the Federal Government' (US), and the General-Purpose AI Code of Practice (EU)), and (iii) whether misalignments between these perspectives warrant closer inspection regarding the viability of governing AI through natural language.
We identify a fragmented literature advancing varying and contradictory claims about what goals system-level instructions can achieve, which we distil into a typology of claims.
Further, we show how divergent claims complicate policy approaches that treat system-level instructions as stable, interpretable control mechanisms. 
We argue that given such misalignments, careful consideration must be given to \textit{prompt governance} approaches. Our findings have broad implications, extending from the immediate LLM policy context %
to the use of natural language as a control mechanism in technical systems more generally.

\end{abstract}



\maketitle

\section{Introduction} \label{sec:intro}

There is currently much attention on generative artificial intelligence (GenAI) models which can be operated through natural language rather than `formal' code~\cite{ouyang_training_2022}. 
Natural language instructions (i.e., `prompts') include end-users' specific queries (e.g., `Draft my homework essay on Taylor Swift'). They also are set by developers and deployers at the system level, where stakeholders implement system-level instructions (e.g., ``You are a helpful assistant'' or ``Avoid generating content involving real public figures''~\cite{claudeSystemPrompts}). 
\textit{System-level instructions} aim to specify the intended conduct, constraints, and priorities that guide language model outputs~\cite{palla_policy-as-prompt_2025, claudeSystemPrompts, microsoftSystemMessage}. These instructions are designed to persist across interactions through training procedures that prioritise some instructions over others~\cite{wallace_instruction_2024, bai2022traininghelpfulharmlessassistant}.

Natural language instructions are inherently more accessible, legible, and modifiable to a wide range of stakeholders. Developers already use them as guardrails and configuration layers for deployed systems~\cite{openai_modelspec, githubGitHubXaiorggrokprompts, zhang_first_2024}. Research literature attributes various intended goals for the function of system-level instructions, from defining model behaviour~\cite{gupta_bias_2023, vachharajani_pjmathematician_2025, iadisernia_prompting_2025}, imposing safety constraints~\cite{microsoftSafetySystem, lin_unlocking_2023, yu_high_2024}, informing domain adaptation~\cite{xu_tool_2023, kagentWriteSystem, openaiPromptEngineering}, to optimising output accuracy~\cite{zhang_agentic_2025, rumiantsau_beyond_2024, wallace_instruction_2024}. Policymakers have begun to treat these same artefacts as potential governance levers: if instruction text can be accessed, inspected, and constrained, then governance might proceed by regulating the language that conditions model behaviour. The United States (\textbf{US}) Executive Order 14139~\cite{EO14319} and European Union (\textbf{EU}) General-Purpose AI Code of Practice~\cite{EUCodePractice2025} both explicitly identify system-level instructions as governable artefacts. While emerging approaches vary, it indicates policy interest to access, evaluate, and constrain language of system-level instructions to inform system behaviour.

Governance of system-level instructions would rely on two aspects:
(a) that accessing system-level instructions reveals what operations the system is \textit{intended} to realise%
; and (b) %
that writing natural language constraints into those instructions causes systems to \textit{function in line} with those constraints~\cite{voxThatApparently, propublicaInsidePrompts, openaiPromptEngineering}. 
Whether system-level instructions achieve these functions in practice is a consequential question. If system-level instructions are treated as governable objects without reliable impacts on behaviour, it risks targeting artefacts that are only imperfect outcome proxies.

Therefore, we \textbf{systematically examine system-level instructions} as governance objects by tracing how their goals, functions, and limitations are \textbf{characterised in \textit{research} and how they are operationalised in \textit{policy}}. On the one hand, research involving system-level instructions is rapidly expanding but heterogeneous; it is therefore difficult to determine which governance-relevant goals are being attributed to system-level instructions, what is treated as evidence, and where uncertainties or contradictions persist~\cite{zhang_iheval_2025, wallace_instruction_2024, guo2026ihchallengetrainingdatasetimprove}. On the other hand, policy instruments increasingly position system-level instructions as governance levers, often without making explicit how prompt text is expected to translate into behavioural outcomes. Our contribution is to make these implicit theories of control explicit, and to identify where prompt-oriented governance aligns with, or departs from, the state of the evidence.
Specifically, we address the following
research questions (\textbf{RQs}):
\begin{itemize}
    \item \textbf{RQ1:} How do researchers discuss goals regarding system-level instructions in academic literature?
    \item \textbf{RQ2:} How are system-level instructions positioned as objects of governance across jurisdictions, and to what extent do policy discussions reflect assumptions about attainability of certain goals?
    \item \textbf{RQ3:} What challenges arise for governing AI through system-level instructions given the states of research and policy discussions internally and their relationship to each other? 
\end{itemize}

Methodologically, we combine %
a systematic literature review that extracts and analyses claims about the goals of system-level instructions (\S\ref{sec:litreview}, \S\ref{sec:litsurveyresults}), and %
a scoping search of policy documents that explicitly address system-level instructions, complemented by two case studies (\S\ref{sec:legalanalysis}). %
We map how researchers discuss what system-level instructions are intended to achieve, characterise how policymakers frame system prompts as governance instruments, and \textbf{surface the assumptions, challenges, and resulting considerations} for governing AI through natural language (\S\ref{sec:discussion}).

In doing so, we find a fragmented research landscape that advances divergent claims about the utility and reliability of system-level instructions, and emerging governance instruments that target these instructions as support for AI system transparency, alignment, and risk mitigation. Often, governance approaches assume system prompts result in stable behavioural control of AI systems, which is not consistently supported by technical evidence. We argue that these mismatches, which can potentially provide a `false sense of control', warrant careful consideration due to implications on instruction, socio-technical, and societal levels.

While this paper focuses on large language models (\textbf{LLMs}), the challenges extend to broader applications as AI systems increasingly incorporate natural-language interfaces. Our findings concern the fundamental utility and reliability of language-governed technology, with implications for accountability, fairness, and transparency across the AI ecosystem.

\section{Understanding System-Level Instructions} \label{sec:background}

Most GenAI systems are operated through text. We focus on LLMs to isolate the effects of natural language for inputs and outputs. This section provides only the technical and policy background needed to situate our analysis; it is not intended as a comprehensive review of prior work, which we instead develop systematically in \S\ref{sec:litreview}. LLMs distinguish between \textit{user prompts}, which express the immediate request for a given interaction, and \textit{system-level instructions}, which set higher-priority constraints and preferences intended to persist across interactions~\cite{wallace_instruction_2024} (see \autoref{fig:promptchain}). %
For this paper, we use \textit{system-level instructions} to refer to natural language prompts that can be issued by human stakeholders and exhibit at least one of three characteristics: \textit{(i)} they persist across conversations (longitudinal), \textit{(ii)} operate at different privilege levels (hierarchical), or \textit{(iii)} function to resolve conflicts between prompts (conflicting). It captures how natural language is used to govern models across multiple points and stakeholders in AI supply chains (\S\ref{sec:background_instruction}). 

An example illustrates the governance intuition of system instructions: A system instruction might state ``avoid generating content involving real public figures''~\cite{claudeSystemPrompts}, while a user asks for a `homework essay about Taylor Swift'. If the system instruction is treated with a higher priority, the model should refuse rather than comply.

\subsection{Instruction Hierarchies and Control Across Supply Chains} \label{sec:background_instruction}

\begin{figure}[ht]
    \centering
    \vspace{-0.7em}
    \includegraphics[width=0.8\linewidth]{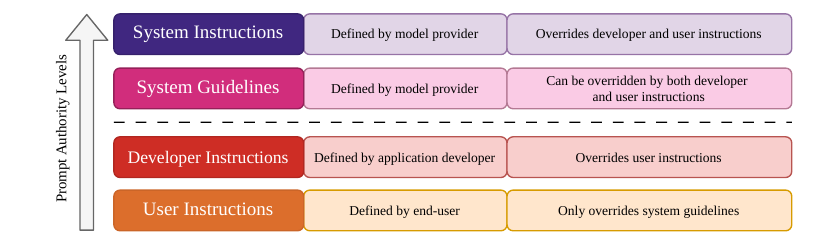}
    \caption{Hierarchical authority levels for different AI instructions (`prompt stack'), with stakeholders and override abilities.}
    \label{fig:promptchain}
\end{figure}

Instruction hierarchies describe how different sources of instruction are prioritised, such as platform-level rules, application developer instructions, and user prompts~\cite{wallace_instruction_2024, guo2026ihchallengetrainingdatasetimprove}.
OpenAI's Model Spec, for example, describes a ``chain of command'' in which higher-level instructions constrain what downstream developers and end-users can do~\cite{openai_modelspec} with control being distributed differently across AI supply chain stakeholders~\cite{10.1145/3715275.3732017, cobbe_account_2023, balayn2024understandingstakeholdersperceptionsneeds, neumann_position_2025, widder_dislocatedaccount}.
Foundation model providers (e.g., OpenAI or Anthropic) typically set baseline rules that ``cannot be overridden by developers or users'', such as safety constraints or restrictions on prohibited content~\cite{ouyang_training_2022, openai_modelspec, claudeSystemPrompts}. Application developers and deployers can add system-level instructions to specialise behaviour for a use case (e.g., a domain assistant, an enterprise policy layer), while remaining limited by higher-level constraints. End users may also set persistent preferences (e.g., style instructions) via platform or API features, but these operate at the lowest priority level and are routinely constrained by upstream instructions~\cite{openaiPlatform, openaiIntroducingGPTs, claudeConfiguringUsing}. The resulting `prompt stack' (see \autoref{fig:promptchain}) means that a model’s outputs can reflect instructions from multiple actors who do not share full visibility into each other’s constraints~\cite{neumann_position_2025, cobbe_account_2023, widder_dislocatedaccount}.

\subsection{System-Level Instructions as Steering Instruments} \label{sec:background_values}

System-level instructions are increasingly treated as instruments of steering as they use natural language to express values and priorities. Nuanced linguistic choices (e.g., `must' versus `should') can signal different degrees of obligation and can affect how models respond~\cite{techpolicyNextAntitrust}.
This linguistic accessibility offers apparent advantages (e.g., interpretability, modifiability), and risks (e.g., ambiguity, inconsistent effectiveness). 
Even high-level instructions like `be helpful' embed normative judgments about what is meant and whose perspective is centred~\cite{zheng_when_2024, korinek2022alignedwhomdirectsocial, jakesch_how_2022}. The design and effect of system instructions are further argued to reflect unexamined assumptions rather than empirically validated approaches~\cite{birhane_values_2022, buyl2025largelanguagemodelsreflect, neumann2026controls}. This motivates scrutiny of whether prompt text can serve as a reliable proxy for behavioural outcomes.

\subsection{Opacity, Transparency, and Malleability} \label{sec:background_pressures}

System-level instructions are used for diverse operational purposes, including setting roles or personas, imposing guardrails, and specifying formatting or tool-use procedures~\cite{microsoftSystemMessage, anthropicGivingClauderole, kagentWriteSystem, openaiPromptEngineering, anthropicEffectiveContext, chappidi_who_2026}. Yet most foundation model providers keep their system prompts confidential, and many models are instructed to refuse revealing them to users~\cite{githubGitHubXaiorggrokprompts, daniel_impact_2024}. Opacity is justified by a mix of concerns, including protecting proprietary information~\cite{openai_modelspec, microsoftProtectedMaterial}, mitigating prompt injection and prompt extraction~\cite{microsoftPromptShields, wallace_instruction_2024}, and preserving safety constraints~\cite{openaiImplementGuardrails}. 

At the same time, pressures for disclosure are increasing. Some companies voluntarily publish system prompts or prompt specifications~\cite{claudeSystemPrompts, githubGitHubXaiorggrokprompts}, and researchers call for transparency to support evaluation and oversight~\cite{neumann_position_2025, widder_dislocatedaccount, norval_dicsbydesign_2022, neumann2026controls}. Some jurisdictions have also begun to require or encourage disclosure of system-level instructions in specific contexts~\cite{EO14319,EUCodePractice2025,DAIEF}.   
The malleability of system-level instructions' complicates research and governance attempts further. Providers can update instructions on short timescales~\cite{githubGitHubXaiorggrokprompts,claudeSystemPrompts} with ad-hoc prompt updates often used as reactive fixes~\cite{nytimesChatGPTWent, wiredOpenAIScrambles, techcrunchTakesGrok, neumann2026controls}.

\subsection{Policy Perspectives on System-Level Instructions} \label{sec:background_policy}
System-level instructions are attracting regulatory attention because they appear to offer an accessible point of intervention. As natural language artefacts, they are legible to non-technical stakeholders and seem to provide a way to 
inspect the behaviour of otherwise complex systems. 
This apparent interpretability distinguishes prompts from other technical levers that are harder to access, interpret, or evaluate~\cite{nannini_explainability_2023, zhou_how_2023, chan_visibility_2024}. Policymakers can regulate the \textit{text}, but it does not guarantee regulation of the \textit{behaviour} that the text is intended to produce.
Prompts are a design lever for AI systems, but their effects remain mediated by training, inference dynamics, and interaction context~\cite{judge_codeisntlaw_2024}. Given the increasing policy focus, the technical conditions under which system-level instructions produce stable, predictable, and verifiable effects is a governance concern, motivating careful engagement with the research evidence.

\section{Literature Review Methodology} \label{sec:litreview}

We conducted a PRISMA systematic literature review to identify and analyse claims about the goals attributed to system-level instructions. 
We use thematic analysis to derive a typology of claims, following established methods~\cite{page_prisma_2021,deck_critical_2024}.

\textbf{Search strategy and sources.} \label{sec:litsurveymethods}
We developed search terms through an initial scoping phase in Google Scholar, used to map terminology, test candidate keywords, and identify relevant databases. This involved screening approximately 600 papers using broad queries and snowballing from key %
papers that define, operationalise, or critically discuss system-level instructions~\cite[i.e.,][]{wallace_instruction_2024, mu_closer_2025, neumann_position_2025, palla_policy-as-prompt_2025, kholkar_policy-as-prompt_2025}. 
We then implemented a structured database search in Scopus, including Scopus Preprints. We restricted the search to 2022-2025, reflecting the emergence of instruction-following paradigms in LLMs~\cite{ouyang_training_2022}. The final search string is reported in~\autoref{app:methodlit}.

\begin{figure}[ht]
    \centering
    \vspace{-1em}
    \includegraphics[width=0.7\linewidth]{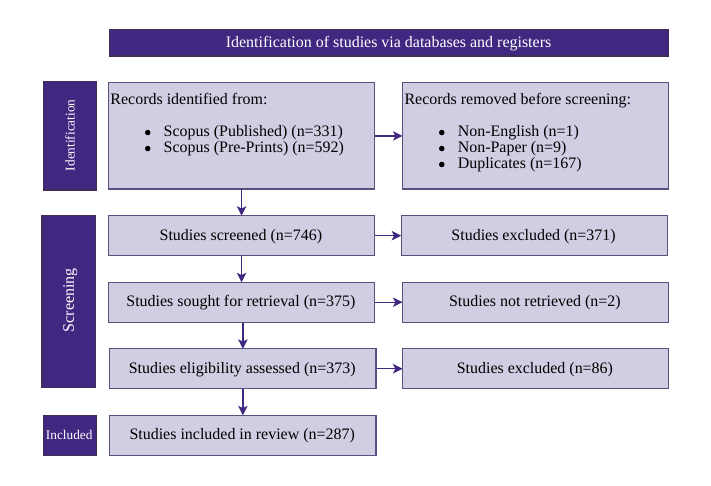}
    \caption{PRISMA flowchart describing our article selection procedure and results.}
    \label{fig:prisma}
    \vspace{-1em}
\end{figure}

\textbf{Screening and eligibility.} Following PRISMA~\cite{page_prisma_2021}, we documented the selection process from 923 identified records to a final corpus of 287 papers (\autoref{fig:prisma}). We removed non-article records (e.g., entire proceedings) and deduplicated entries across published and preprint sources. Abstract screening (746 records) excluded papers that (i) did not engage with our definition of system-level instructions (\S\ref{sec:background}); (ii) used ``prompt policy'' only in mathematical reinforcement learning; or (iii) studied non-text-only systems (e.g., vision-language models), to isolate text-to-text instruction effects. 

\textbf{Claim extraction.} We conducted full-text review of 373 papers. Following methods in \cite{deck_critical_2024}, we heuristically scanned each paper for explicit claims and extracted \textit{unique statements about the intents or outcomes of system-level instructions}. We excluded papers that restated prior work without adding a distinct claim, and papers that were too vague to attribute a specific goal. Claims were recorded and collected in Zotero~\cite{zotero}, then exported, and quantitatively analysed with custom-made Python~scripts.

\textbf{Thematic coding and typology construction.} \label{sec:litsurveythemanalysis}
We analysed extracted claims using inductive thematic analysis~\cite{clarke_thematic_2017}. Each claim received (i) a goal code capturing the underlying intent (e.g., `set domain context') and (ii) a stance label indicating whether the claim describes system-level instructions as \textit{furthering} or \textit{hindering} that goal. We then grouped goal codes into higher-level categories, resulting in an eight-category typology (\autoref{fig:typology_goals}). Additional detail on coding decisions and category assignment is provided 
in~\autoref{app:coding}, and paper mappings are listed in~\autoref{app:corpusrefs}.

\section{A Typology of System-Level Instruction Goals} \label{sec:litsurveyresults}

\begin{figure}[ht]
    \centering
    \vspace{-0.5em}
    \includegraphics[width=0.8\linewidth]{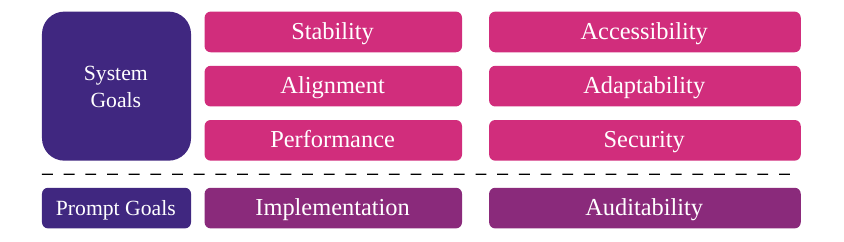}
    \caption{Typology of system-level instruction goals derived through thematic analysis of claims.}
    \label{fig:typology_goals}
    \vspace{-1.5em}
\end{figure}

Through thematic analysis of claims in our corpus, we identified eight categories of goals related to system-level instructions. These goals divide into two types based on their target of intervention: \textit{system goals} and \textit{prompt goals}~(\autoref{fig:typology_goals}). %
\textbf{System goals} refer to instructions that target the behaviour of the AI system itself. Researchers make claims concerning whether system-level instructions can shape model behaviour and change their outcomes. In contrast, \textbf{prompt goals} target system-level instructions themselves as specific artifacts. Researchers make claims about their creation, implementation, or optimisation, as well as the processes of their engineering. %

Some claims addressed multiple goals; in these cases we determined a primary category through collaborative discussions (see~\autoref{app:coding}). Following, we detail how researchers discuss each goal through supportive and critical claims, i.e., claims that support that the goal could be advanced (\textit{furthered}) or that are critical of the goal being advanced (\textit{hindered}). We provide a table of exemplary quotes for all eight categories in~\autoref{app:exclaims}, \autoref{tab:claim_examples}.%
\begin{pinkbox}
    \subsection{System Goal: Alignment} \label{sec:sysgoalalign} 
    \textit{Align LLMs with the goals and values of stakeholders, including developer, deployer, user, and policymaker interests.} 
\end{pinkbox} %

\noindent
A central question in the literature is whether system-level instructions can shape \textit{system behaviour}
in ways that reflect normative commitments, ranging from broad principles embedded during training to specific operational constraints such as content moderation policies. 
Researchers discuss alignment both as broad stakeholder goals (e.g., user safety, regulatory compliance, enterprise policy adherence) and as more concrete, testable behaviours.

\textit{\textbf{Further.}} %
A range of studies report that system level instructions can meaningfully align system behaviour under certain conditions, e.g., explicit value specifications can reduce the frequency of unsafe outputs~\cite{greenblatt_alignment_2024, alan_improving_2025, ermolaeva_how_2024, sovrano_can_2025, ben-zion_detecting_2025}, or priority specifications can ensure models resolve conflicts between competing values~\cite{leidinger_how_2024, liu_generative_2025}. Other work highlights practical benefit of runtime alignment, i.e., prompts allow deployers to adjust behavioural constraints without retraining~\cite{le_conversational_2025, kucharavy_adapting_2024, ahlgren_assisting_2025}. It can reduce bias or boost diversity in accordance with instructed values~\cite{wang_assessing_2025, kumar_can_2025, jeong_system_2025, yusuf_rag-based_2024}, enforce audience-specific content moderation~\cite{khoo_minorbench_2025}, or enterprise policies~\cite{kholkar_policy-as-prompt_2025, neelou_a2as_2025, wallace_instruction_2024}.

\textit{\textbf{Hinder.}} System-level instructions can be used to induce harmful or otherwise misaligned behaviour, including by configuring pursuance of malicious objectives~\cite{zhan_malicious_2025}. More generally, stating values in natural language does not guarantee models will behave accordingly, nor that conflicts will be resolved as intended~\cite{leidinger_how_2024}. Several studies suggest that models align more reliably with pre-defined value systems (e.g., `helpfulness') than user-defined ones~\cite{leidinger_how_2024, schlatter_shutdown_2025, chiu_dailydilemmas_2025}, and that other methods influence what instruction language can be reliably followed~\cite{kumar_can_2025, khoo_minorbench_2025, bernstein_trust_2025, sun_beyond_2024}. Models remain operationally unsafe for some use cases despite explicit safety directives~\cite{lei_offtopiceval_2025}, exhibit alignment-faking that resists correction~\cite{greenblatt_alignment_2024}, and can introduce or amplify biases through system prompts~\cite{neumann_position_2025, kumar_can_2025}. Expectations of authors of system-level instructions can also diverge from model behaviour in practice, showing unreliable  alignment~\cite{karny_neural_2025}.

\begin{pinkbox} 
    \subsection{System Goal: Accessibility} \label{sec:sysgoalaccess}
    \textit{Make model control accessible to stakeholders through reliable natural language interfaces and rapid editing.}
\end{pinkbox} %

\noindent 
The literature frames accessibility as the extent to which system-level instructions allow stakeholders to steer, reconfigure, or understand the underlying systems rather than using specialised technical interventions. System-level instructions are perceived as `easy' instruments to \textit{configure} systems as researchers discuss advantages of them being less resource-intensive~\cite{kucharavy_adapting_2024, ahlgren_assisting_2025, zhang_agentic_2025} and easy to \textit{deploy} via APIs and chat templates~\cite{heimburg_complementor_2025, neelou_a2as_2025, zhang_autonomous_2023, zhang_first_2024}.

\textit{\textbf{Further.}} Several studies characterise system-level instructions as accessible interventions: modifying models through these can be ``almost seamless''~\cite{hazan_security_2025}, and can allow developers to ``instantly govern model behaviour by editing''
prompts rather than more technical design choices~\cite{lee_data-model_2025}. Researchers claim the natural language interfaces enable immediate behavioural reconfiguration~\cite{furmakiewicz_design_2024, hofmann_beyond_2025, xu_tool_2023, hui_span_2024} and may enhance transparency or explainability by making intended constraints and roles more legible to human stakeholders~\cite{ramos_transparent_2024, le_conversational_2025}. %

\textit{\textbf{Hinder.}} Some papers caution that the ease of accessing prompts does not necessarily translate into increased understanding of their effects. Even where instruction text is readable, model behaviour can be difficult to predict, such that ``users are essentially operating blind''~\cite{karny_neural_2025}. Evaluation of prompt text often relies on trial-and-error \cite{lee_large_2025, furmakiewicz_design_2024, palla_policy-as-prompt_2025}. This is reflected in the prominence of prompt engineering as a largely experience-driven practice with variable outcomes~\cite{wang_rnr_2024, hazan_security_2025, nguyen_software_2025, modi_assessing_2025}. Moreover, automatically-generated prompts can be difficult to interpret~\cite{mcginness_large_2025, challagundla_si-agent_2025} or ``remarkably different from anything a human practitioner would be likely to generate''~\cite{battle_unreasonable_2024}, reducing the accessibility benefits for human oversight. %

\begin{pinkbox}
    \subsection{System Goal: Adaptability} \label{sec:sysgoaladapt} 
    \textit{Adapt models to different domains, contexts, behaviours, LLM personas, or purposes through runtime configuration.}
\end{pinkbox} %

\noindent
Adaptability is framed as the capacity to reconfigure model behaviour through system-level instructions for domain-specific applications (e.g., specialised assistants), policy or organisational requirements, and interactional styles (personas/roles). In this sense, prompts are a mechanism for translating contextual requirements into behavioural constraints.

\textit{\textbf{Further.}} Many studies report that system-level instructions can support practical forms of adaptation across tasks and settings. This includes configuring outputs to reflect user preferences or interaction styles~\cite{gallego_configurable_2024}, encoding policy guidelines~\cite{kholkar_policy-as-prompt_2025, lichkovski_eu-agent-bench_2025}, and domain adaptation~\cite{zheng_reasoning_2025}. Persona prompting, i.e., giving the AI a persona to embody like `a helpful assistant', is frequently presented as a core use case: system-level instructions are used to specify roles, goals, and behavioural expectations~\cite{NEURIPS2024_de7b9910, lu_rolemrc_2025, iadisernia_prompting_2025}. Researchers report using system-level instructions to configure specialised system types, such as autonomous agents~\cite{zhang_personaagent_2025, shen_optimizing_2025, panda_causal_2025}, or LLM-as-a-Judge evaluation systems~\cite{kirichenko_abstentionbench_2025, chiang_large_2024, ben-zion_detecting_2025}. Some treat it as fundamental to system design, claiming the ``key strength of this design lies in the fact that the assistant's behaviour is governed entirely by the system prompt''~\cite{le_conversational_2025}.

\textit{\textbf{Hinder.}} Researchers emphasise that prompt-based adaptation is not uniformly reliable.
Models do not enforce all stated rules equally~\cite{chiang_large_2024, scheurer_large_2024}, particularly `forgetting' instructions in lengthy~\cite{qiu_training_2024, chiang_why_2025, almasi_alignment_2025}, or conflicting~\cite{zou_eifbench_2025, qiu_training_2024, jaroslawicz_how_2025} contexts.
Studies report that models lack generalisation when presented with unfamiliar domains~\cite{liu_analyzing_2025, qu_beyond_2025}, taxonomies~\cite{li_gspr_2025}, or tasks~\cite{lee_aligning_2024, kumar_can_2025, xu_tool_2023}. Persona prompting shows conflicting results: roles can be ``brittle''~\cite{maiya_open_2025}, have minimal influence on response content~\cite{li_effects_2024, kumar_can_2025}, or their effect ``might largely be random''~\cite{zheng_when_2024}. 

\begin{pinkbox}
    \subsection{System Goal: Performance} \label{sec:sysgoalperf} 
    \textit{Improve model performance on tasks, benchmarks, or interaction quality through varying methods.}
\end{pinkbox} %

\noindent
Researchers make claims about system-level instructions boosting system performance across various metrics~\cite{lee_large_2025, liang_integrating_2025, chantangphol_finmind-y-me_2025}, from task accuracy~\cite{rumiantsau_beyond_2024, zeng_evaluating_2024, bernstein_trust_2025, liu_oraplan-sql_2025} to response quality~\cite{bo_disclosures_2024, lee_data-model_2025}. Performance improvements are discussed both as stand-alone benefits and as advantages over alternative optimisation methods~\cite{zhang_autonomous_2023, zhao_adversarial_2024}.

\textit{\textbf{Further.}} Many studies report that well-designed system-level instructions can improve performance in targeted settings. Reported gains include increases in task accuracy~\cite{lee_large_2025, costa_instruction-level_2025, liu_oraplan-sql_2025}, with domain-specific instruction design producing measurable enhancements~\citep{liang_integrating_2025}. Several papers argue that greater specificity or longer prompts can improve performance~\cite{yao_leveraging_2024, yuan_dmt-rolebench_2025, braun_hidden_2025}, enabling ``faster and cheaper customization than fine-tuning while delivering strong specialized performance''~\cite{levin_has_2025}. Prompting can achieve better accuracy than data retrieval methods~\cite{adebimpe_sbash_2025}.

\textit{\textbf{Hinder.}} Other work suggests performance gains are uneven and context-dependent. Fine-tuning can outperform system prompting~\cite{marquardt_fine-tuning_2025, shi_llm-augmented_2025}, and some studies report no measurable effect of system prompts on performance~\cite{li_effects_2024}. Practical constraints further complicate performance claims, as long or highly detailed prompts can introduce latency, throughput bottlenecks, and higher costs~\cite{zhu_relayattention_2024, chen_pitfalls_2025, liang_integrating_2025}. Persona specifications are discussed as a potential source of performance degradation~\cite{zheng_when_2024}, and as unstable for achieving particular accuracy requirements~\cite{kumar_can_2025}.

\begin{pinkbox}
    \subsection{System Goal: Stability} \label{sec:sysgoalstabl} 
    \textit{Maintain consistent and predictable model behaviour and instruction following overall.}
\end{pinkbox} %

\noindent
Stability concerns whether system-level instructions produce sufficiently consistent and reliable behaviour. %

\textit{\textbf{Further.}} Several papers argue that instruction hierarchies, multi-layered approaches, and explicit prioritisation can help maintain consistent behaviour~\cite{zhang_iheval_2025, wallace_instruction_2024, wang_rnr_2024, das_system_2025, neelou_a2as_2025, krishna_d-rex_2025, zheng_reasoning_2025, liu_analyzing_2025, cohen_information_2025}. %
These effects can generalise across domain variations or interaction length~\cite{greenblatt_alignment_2024, kim_debiasing_2025}, while possibly ``maintaining or enhancing the model's [...] capabilities'' \cite{wu_automedprompt_2025}.

\textit{\textbf{Hinder.}} However, system-level instructions can also struggle to achieve stable effects~\cite{marquardt_fine-tuning_2025, leidinger_how_2024, jaroslawicz_how_2025}. Researchers show system prompts only achieve moderate accuracy when resolving instruction conflicts~\cite{zhang_iheval_2025, qin_sysbench_2024}, sometimes even following lower-priority instructions more reliably than instructions that are indicated as higher-priority%
~\cite{schlatter_shutdown_2025}. Instructions can be unstable over multi-turn interactions~\cite{li_effects_2024, chen_parapo_2025}, and constraints can be violated despite explicit directives~\cite{qin_sysbench_2024}, e.g., `do not act against these instructions'. Outputs can be significantly less stable with more or conflicting guidelines \cite{zhang_iheval_2025, zou_eifbench_2025}, with changed instruction orders~\cite{chen_pitfalls_2025}, and specific social cues placed in user inputs~\cite{zeng_who_2025}. Overall, instructions can introduce various instabilities~\cite{zhang_iheval_2025, zou_is_2024, zhan_malicious_2025}.%

\begin{pinkbox}
    \subsection{System Goal: Security} \label{sec:sysgoalsecur}
    \textit{Maintain or improve system security with (or despite) system-level instructions.}
\end{pinkbox} %

\noindent
Research discusses the impact of system-level instructions on security through three lenses: as attack vectors used by adversaries to compromise models~\cite{maloyan_adversarial_2025, zhang_enja_2024, chao_jailbreakbench_2024}, as defence mechanisms protecting against malicious inputs~\cite{guo_review_2024, debenedetti_dataset_2024, xie_defending_2023}, and as vulnerable assets requiring protection from extraction or manipulation~\cite{krishna_d-rex_2025, gakh_enhancing_2025, pape_prompt_2025}.

\textit{\textbf{Further.}} Several papers frame system-level instructions as a practical defence mechanism, instructing a model to follow safety policies, e.g., `defend against jailbreak attacks'. Defensive prompts can achieve high extraction protection~\cite{zhuang_proxyprompt_2025} and low prompt leakage~\cite{gakh_enhancing_2025}. Such guardrails can reduce jailbreak success and block significant portions of malicious inputs~\cite{xie_defending_2023, kholkar_policy-as-prompt_2025}, although trade-offs between security and performance are noted~\cite{pfister_gandalf_2025}.

\textit{\textbf{Hinder.}} System-level instructions are repeatedly described as an enabling factor for attacks~\cite{zhang_agent_2025, salem_maatphor_2023, li_system_2025}, as system jailbreaks can require ``nothing more than a specific system prompt''~\cite{hagendorff_large_2025, li_system_2025}. Because system prompts can include ``configuration details, user roles, and operational instructions''~\cite{das_system_2025}, they are also treated as valuable targets for adversaries, and prompt disclosure could directly undermine system integrity~\cite{yang_promptcos_2025, wu_slip_2025, ogundoyin_unsafe_2025,chiang_why_2025}. Researchers report consistently high system prompt extraction success across methodologies and models \cite{hui_span_2024, wang_ip_2025, nie_leakagent_2025}, with further technical boundaries like filters failing to prevent substantial extraction \cite{zhang_extracting_2024}.

\smallskip
    \textbf{These six goals were oriented to the system operations, while the following two goals relate to the specific system instructions.}

\begin{purplebox}
    \subsection{Prompt Goal: Implementation} \label{sec:progoalimpl} 
    \textit{Determine effective methods for authoring, structuring, deploying, and maintaining system-level instructions as technical artefacts themselves.}%
\end{purplebox} %

\noindent
Implementation research focuses on how system-level instructions should be produced and operationalised in practice. Common questions include whether prompts should be human-written or automatically generated, how prompts should be structured and maintained over time, and whether prompts should be static or update dynamically during interaction. Choices are discussed regarding effectiveness, security, and accountability.

\textit{\textbf{Further.}} Some studies argue that automated or optimisation-based prompt generation can produce prompts that are production-ready prompts and improve performance \cite{costa_instruction-level_2025, wang_rnr_2024, battle_unreasonable_2024}. Other work emphasises the value of human authoring, though perceived as more resource-intensive~\cite{challagundla_si-agent_2025}, particularly where domain expertise and interpretability are important~\cite{levin_has_2025}. Regardless of generation, researchers stress the benefits of iterative implementation~\cite{zhang_first_2024, lee_large_2025}: ``A good system prompt will only be arrived at through iterative trial and error'' \cite{furmakiewicz_design_2024}.
Regarding prompt length, some researchers claim that more specific prompts work best~\cite{yao_leveraging_2024}, others state that shorter prompts deliver better results~\cite{mu_closer_2025, duan_hierarchical_2025, shayegani_measurement_2025}. Additionally, some researchers argue continually optimised~\cite{gallego_metasc_2025, costa_instruction-level_2025} or re-injected prompts~\cite{neelou_a2as_2025, yan_refutebench_2024} are superior implementations.

\textit{\textbf{Hinder.}} Several papers identify risks associated with prompt optimisation and ad hoc engineering. Automatically generated prompts may drift from intended meanings, increasing risk of unexpected behaviour~\cite{sharma_sysformer_2025}, and extraction vulnerabilities~\cite{peng_repeatleakage_2025}. Human-authored prompts are not ideal implementation methods either as they do not have clear paths toward systematic improvement~\cite{challagundla_si-agent_2025, akaybicen_machine_2024, almasi_alignment_2025}. Researchers warn that ``excessive reliance raises the risk of manipulation by malicious actors''~\cite{chan_detection_2024}, which is worsened by a lack of security practices~\cite{zhang_first_2024, pham_cain_2025, zhang_effective_2024}.

\begin{purplebox}   
    \subsection{Prompt Goal: Auditability} \label{sec:progoalaudit} 
    \textit{Make system-level instructions auditable, i.e., visible, traceable, and included in system evaluations.}
\end{purplebox} %

\noindent
Auditability concerns whether system-level instructions can themselves be meaningfully overseen and evaluated in practice, and under what conditions. The literature identifies core issues as the transparency of instructions~\cite{takeshita_gengo_2025, kempny_qualipilot_2025, neumann_position_2025}, transferability of instruction effects~\cite{agarwal_prompt_2024, peng_repeatleakage_2025}, and specific oversight and evaluation methods~\cite{navneet_rethinking_2025, doudkin_spark_2025}.

\textit{\textbf{Further.}} Some work proposes concrete mechanisms to support prompt auditing, including standardised evaluation frameworks~\cite{kang_doppelganger_2025, cobbina_where_2025}, expert oversight of prompt ecosystems~\cite{hazan_security_2025, ogundoyin_unsafe_2025, cobbina_where_2025, muti_customizing_2024}, and documentation practices that enable prompts to be traced across decisions made%
\cite{palla_policy-as-prompt_2025}. If system prompt effects transfer robustly across variations, models, or domains, single evaluations could inform broader assessments and enable ecosystem-wide evaluation schemes.

\textit{\textbf{Hinder.}} The literature repeatedly notes barriers to both transferability~\cite{duan_hierarchical_2025, almasi_alignment_2025, levin_has_2025, daniel_impact_2024} and transparency~\cite{neumann_position_2025, mcginness_large_2025}. Transferability is often described as conditional, with prompt effects varying across models, domains, and small changes in phrasing or structure~\cite{amjad_agentic_2025, palla_policy-as-prompt_2025, zhang_automated_2025}, limiting the generalisability of audit findings~\cite{weesep_exploring_2025, thelwall_evaluating_2025}. Limited transparency prevents open evaluation practices and design of alternatives with or for proprietary models~\cite{neumann_position_2025, mcginness_large_2025}. Disclosure is treated as a goal for prompt transparency~\cite{palla_policy-as-prompt_2025, takeshita_gengo_2025, chiang_large_2024} but is made less effective through the practices of frequent prompt iterations, hindering reconstruction of active versions at a specific point %
~\cite{doudkin_spark_2025, tuggener_role-playing_2024, zhang_sprig_2024}, and incomplete specifications, which can also undermine fixed audit regimes~\cite{neumann_position_2025, das_system_2025, wang_illusion_2025}.
\bigskip

\textbf{Overall}, the research landscape suggests that while system-level instructions have several functions and goals, there are considerable uncertainties and contradictions in the evidence. Nevertheless, we examine how governance frameworks operationalise system-level instructions as governable artefacts, and what assumptions~this~embeds. 

\section{Legal Aspects of System Prompt Governance} \label{sec:legalanalysis}

We conducted a broad and practice-oriented search for policy and governance documents that explicitly address system-level instructions. System-level instructions appear as an object of governance in only a small number of documents, as it is a new technological artifact, with explicit references only emerging across multiple jurisdictions and settings. We identified eight relevant documents. Our search methodology is detailed in~\autoref{app:methodlegal}. %

The aim of the following analysis is not to offer a doctrinal interpretation of legal obligations, but identify how system-level instructions are considered in governance discussions as a means for uncovering \textit{assumptions} about system-level instructions as objects of governance. We therefore read these documents as articulating implicit theories of control: what regulators expect to achieve with and through system-level instructions.

\subsection{Current Approaches} \label{sec:legalanalysisapproaches}

We briefly describe the retained documents here, before turning to two case studies. Despite several documents mentioning system-level instructions, these are the only ones %
that to our knowledge substantively engage with them as distinct governable artefacts.
The Australian Digital Transformation Agency is conducting a trial of LLMs evaluating procurement applications and will ``use change controls for its AI material (models, prompts and data sets) to document the evaluation methodology for each panel, ensuring consistent, replicable and defensive results''~\cite{DTA2024}. 
The \textit{Artificial Intelligence Playbook for the UK Government} (Feb' 2025) \cite{UKPlaybook2025} and the \textit{Code of Practice for the Cyber Security of AI} (Jan' 2025) \cite{UKCyber2025} provide guidance on security risks in LLMs posed by jailbreaking system-level instructions. Recently, the UK updated its \textit{Data and AI Ethics Framework} (Dec' 2025) that guides public sector AI development. It explains projects should be transparent, including ``document[ing] information about models, data sets and system prompts, and mak[ing] sure operators understand how the system works in different situations''~\cite{DAIEF}. Singapore’s \textit{Starter Kit for Safety Testing of LLM-Based Applications} (Draft for Public Consultation, June 2025) proposes testing to assess how system prompts affect behaviours such as data disclosure and over-filtering~\cite{imda2025}. The EU’s approach defines system prompts in a broader ``model specification'' intended to be disclosed to evaluators~\cite{EUCodePractice2025}. The US approaches to federal procurement of LLMs explicitly discusses system prompt disclosure as a permissible means of evidencing alignment with procurement-facing principles~\cite{EO14319,Memorandum2025}.

\subsection{Case Studies: System Prompt Governance} \label{sec:legalanalysiscasestudies}

We selected two case studies that are distinctly explicit in their references to system-level instructions, and representative of two distinct regulatory styles. Both frame system-level instructions as legible artefacts that can support governance objectives, but operationalise this idea through different mechanisms for different~aims.

\subsubsection{US Federal Procurement} \label{sec:legalanalysiscasestudyUS}

Our analysis focuses on two primary documents: Executive Order (\textbf{EO})~14319, \textit{Preventing Woke AI in the Federal Government} (July 2025) \cite{EO14319}, and the Office of Management and Budget (\textbf{OMB}) memorandum, \textit{Increasing Public Trust in Artificial Intelligence Through Unbiased AI Principles} (Dec' 2025) \cite{Memorandum2025}. Both define procurement-relevant requirements for LLM vendors and establish a federal approach to transparency, neutrality, and accountability in the procurement and deployment of GenAI systems. 
The EO~14319 established that US federal agencies shall only procure LLMs developed in accordance with two core principles, the `Unbiased AI Principles'~\cite[Sec. 3]{EO14319}:

\begin{list}{}{\footnotesize\leftmargin=1em\rightmargin=1em}
\item[]
(a)~\textit{Truth-seeking}. LLMs shall be truthful in responding to user prompts seeking factual information or analysis. LLMs shall prioritize historical accuracy, scientific inquiry, and objectivity, and shall acknowledge uncertainty where reliable information is incomplete or contradictory.

(b)~\textit{Ideological Neutrality.} LLMs shall be neutral, nonpartisan tools that do not manipulate responses in favor of ideological dogmas such as DEI. Developers shall not intentionally encode partisan or ideological judgments into an LLM's outputs unless those judgments are prompted by or otherwise readily accessible to the end user.
\end{list}

Notably, the EO specifies that vendors may demonstrate compliance with the ideological neutrality principle ``through disclosure of the LLM’s \textit{system prompt}, specifications, evaluations, or other relevant documentation'' \cite[Sec. 4]{EO14319}. In December 2025, the OMB memorandum translated the EO into implementation guidance for federal agencies \cite{Memorandum2025}. 
It defines a minimum threshold for LLM transparency that is primarily concerned with the provision of vendor model, system and/or data cards \cite[1(A)(i)]{Memorandum2025}. However, system prompts are usually not reported in model or system cards \cite{Mitchell_modelcards_2019, deepmindModelCards, huggingfaceModelCards}.
Agencies are advised that, depending on the planned use of an LLM, they may request additional information directly relevant to the Unbiased AI Principles. This is categorised by (i) pre-training and post-training activities, (ii) model evaluations, (iii) enterprise-level controls, and (iv) third party modifications. 

\textbf{System prompts are outlined in all of these but model evaluations.}
First, the memorandum treats system-level instructions as part of the model’s development and post-training context, i.e., natural-language guidance that shapes how the model responds to user queries~\cite[1(B)(i)(B)]{Memorandum2025}. Second, it frames them as enterprise governance controls that can be layered on top of a base model’s prompt stack or combined with other output-shaping mechanisms such as content filters~\cite[1(B)(iii)(A)]{Memorandum2025}. Third, it includes system-level instructions among downstream or third-party modifications that vendors may need to disclose when they apply additional controls to steer a model they did not originally develop~\cite[1(B)(iv)(A)]{Memorandum2025}.

The OMB framework explicitly acknowledges a layered hierarchy of system prompts across foundation model developers, enterprise-level system prompts, and third party vendors (see \S\ref{sec:background}, \cite{neumann_position_2025}). 
Therefore, the US federal procurement framework treats system prompts as an optional transparency artifact (\textit{auditability prompt goal}).

However, there are limitations to the way the OMB framework seeks to govern system-level instructions. The framework does not clearly require performance evaluation, nor how they should be empirically assessed. This is in contrast to the guidance on output-level bias evaluations. Importantly, the exclusion of system prompts from the ``model evaluations'' section \textbf{reflects an implicit assumption that inspecting prompt language is sufficient}. In effect, the framework aims for \textit{auditability prompt goal} at the level of artefact disclosure, without clearly operationalising the conditions under which disclosed prompts can be treated as reliable levers towards the \textit{alignment or adaptability system goal}. 
System prompts are treated as static artifacts rather than evolving mechanisms, omitting prompt iteration or layering documentation (\textit{implementation and auditability prompt goal}). %

\subsubsection{EU AI Act Compliance} \label{sec:legalanalysiscasestudyEU}

On 10 July 2025, the EU published the \textit{General-Purpose AI Code of Practice} (\textbf{Code}). It is a voluntary tool designed to help providers of general-purpose AI models follow the AI Act's obligations. While the Code itself is not binding law, it is explicitly designed to operationalise governance expectations through documentation and evaluation practices. It therefore provides a useful insight into how EU governance frameworks conceptualise system-level instructions as governable artefacts, particularly in the context of systematic risk. 

The Code’s Safety and Security chapter treats system-level instructions as part of the model specification and evaluation regime for advanced models presenting systematic risk~\cite[Art.~55]{AIAct},~\cite[Measure~7.1, Appendix~3.4]{EUCodePractice2025}. It requires a ``model specification'' that states the model’s intended principles, their prioritisation, refusal topics, and the system prompt~\cite{EUCodePractice2025}. Evaluation teams should receive this specification (including the system prompt) alongside other relevant materials~\cite{EUCodePractice2025}.

The Code clearly reflects several instruction goals from our typology.
First, the \textit{auditability prompt goal}: %
documenting and scrutinising system-level instructions as legible artefacts. Second, the \textit{accessibility system goal}: disclosing the system instruction to evaluators to make the system more accessible.
It also reflects typology goals in treating the system prompt as a meaningful representation of intended behaviour (\textit{adaptability system goal}) and priority-setting (\textit{alignment system goal}), but does not specify assessment of stable prompt effects (\textit{stability system goal}), despite diverging claims about stability (see~\S\ref{sec:litsurveyresults}).

The Code does not operationalise the use of multiple layers of system-level instructions across the supply chain (\textit{implementation prompt goal})%
, including base model instructions, deployer constraints, and application-level prompts \cite{openai_modelspec, neumann_position_2025, cobbe_account_2023}. It does not require prompt versioning, change logs, or triggers for re-evaluation when system prompts are updated~(\textit{auditability prompt goal}). Without such mechanisms, disclosure can quickly become outdated \cite{palla_policy-as-prompt_2025}. Overall, the EU treats system prompts as part of model specification and evaluation practice. 

\subsubsection{Comparison between case studies} \label{sec:legalanalysiscomparison}

Across both case studies, system-level instructions are treated as governance-relevant not only because they are written in natural language and can be disclosed, inspected, and revised, but because they are presumed to shape system
behaviour: regulators treat prompt language as a proxy for models performance. This shared premise activates the \textit{accessibility system goal} and \textit{auditability prompt goal}. However, the case studies diverge in what they operationalise and what they leave implicit; both show misalignments with the research~literature.

\textbf{Shared emphasis on disclosure as governance.} Both frameworks position system prompts as legible artefacts to support oversight. The US frames disclosure as permissible evidence to comply with procurement principles~\cite{Memorandum2025}. The EU defines the system prompt within the model specifications provided to evaluators~\cite{EUCodePractice2025}. 

\textbf{Divergence on auditability through evaluation.} The EU Code more directly operationalises system-level instructions as artefacts for evaluation practice by providing it to model evaluation teams~\cite{EUCodePractice2025}. It frames prompt disclosure as one aspect of evaluation, aiding interpretation against stated priorities. 
The US framework, by contrast, positions system prompts as optional transparency artefacts and does not specify how their behavioural effects should be empirically assessed, despite more attention to output-level evaluations in other areas~\cite{Memorandum2025}. %

\textbf{Different assumptions about alignment and stability.} The US framework more strongly implies that writing high-level normative commitments into system-level instructions can support the \textit{alignment system goal} with articulated values~\cite{EO14319}. It also implicitly relies on the \textit{stability and adaptability system goals}, i.e., value statements expressed in natural language translate into sufficiently consistent behaviour across contexts. The EU Code is less value-prescriptive, but still assumes the system prompt is a meaningful and usable representation of intended behaviour and priority-setting for evaluators~\cite{EUCodePractice2025}. These assumptions do not align with claims in research literature that describe effects as context-dependent, sensitive to phrasing and ordering, and vulnerable to interaction effects across multi-turn conversations and layered instructions. %

These case studies illustrate a policy move to govern behaviour via the language that governs GenAI. Our analysis suggests this is only plausible if disclosure of prompt language is a starting point for governance and evaluation rather than a substitute for behavioural evidence, and if governance frameworks ensure that they \textbf{address layered, changeable, and security-sensitive system-level instructions}.

\section{Discussion} \label{sec:discussion} %

Our systematic literature review reveals fragmented claims across the typology of system-instruction goals. 
Policy approaches likewise diverge in which goals are emphasised {and operationalised} across jurisdictions.
Despite these divergences, two core assumptions can be made out: that stakeholders can (i) infer intent from instruction text, and (ii) accurately predict model behaviour from those instructions. Our findings reveal that neither assumption reliably holds. Prompt-based governance built on these assumptions risk creating a \textbf{false sense of control}, where policymakers believe that they have addressed AI system risks when fundamental questions remain unresolved.

We synthesise governance implications at three levels: the instruction level (where human and machine readings of the same language diverge), the socio-technical level (where instructions are embedded in complex, multi-actor systems), and the societal level (where governance frameworks rest on both preceding levels).

\subsection{Instruction-Level Implications} \label{sec:discnarrow}

As system-level instructions are operationalised as specific artifacts into socio-technical systems, it is important to consider the effects of language being understood differently by humans and LLMs.
Natural language governance of AI systems appears \textbf{deceptively straightforward}. Controlling complex systems through natural language seems practical, as stakeholders can write instructions, update them iteratively, and deploy them without specialised expertise or infrastructure changes. This accessibility and ease of deployment makes system-level instructions appear to be quick fixes for persistent challenges like bias, safety, and performance~\cite{wang_assessing_2025, guo_review_2024, lee_large_2025, naik_agentmisalignment_2025}.
However, the very accessibility that makes prompts attractive also introduces governance risks.

The ease of prompt intervention may \textbf{encourage expansion in scope}. Once system-level instructions are treated as a viable control mechanism for one objective, it encourages adoption for additional objectives, including those not, or only conditionally, suitable for prompt-based control~\cite{tuggener_role-playing_2024, vachharajani_transformer_2025}. A prompt-centric governance approach may therefore face \textbf{``function creep''}~\cite{Koops02012021_functioncreep}, where it applies to an expanding set of governance tasks, and without corresponding evidence that instructions can reliably deliver each task. 
Through availability in user interfaces~\cite{chromePromptChrome, ollamaGenerateResponse}, this functionality is already deployed at scale and usage will likely increase, e.g., in agent~\cite{githubGitHubTallesborgesagenticsystemprompts, openaiNeuAgentKit} and multi-agent frameworks~\cite{zhou2025multiagentdesignoptimizingagents}. The tendency to use instructions as general purpose mechanisms (\S\ref{sec:sysgoaladapt}) raises clear concerns about value imposition or censorship~\cite{joinrebootAlignmentCensorship}.

Moreover, linguistic accessibility risks \textbf{importing human interpretive assumptions} into machine governance. Regulators necessarily interpret natural language prompts using social and institutional understandings of meaning, obligation, and intent. Navigating the complex interpretation of legal and policy language is not comparable to the interpretation of machine processing of natural language~\cite{janecek_can_2025, Chesterman_Bennett_Moses_Pagallo_2023}. Models operationalise instruction text through several layers of statistical pattern matching shaped by pre-training, post-training, and retrieval regimes~\cite{varshney2019pretrainedaimodelsperformativity, lai-etal-2025-survey-posttraining}, and outputs are sensitive to phrasing and context~\cite{su2025singlecharactermakebreak, goloujeh_isitaiorisitme_2024, ismithdeen2025promptceptionsensitivelargemultimodal}. Translating instructions into GenAI systems requires understanding attention mechanisms, biases, statistical irregularities and stochasticity, and opaque (non-)compliance~\cite{vaswani2023attentionneed,tang_ai_2024, dominguez_hernandez_mapping_2024, hopkins_recourse_2025}.

In all, we argue that these gaps can produce a form of \textbf{compliance illusion}. A system prompt may read as aligned with a governance principle while failing to yield stable adaptation across contexts, models or multi-turn interactions, as reflected in the mixed evidence base identified in our review (\S\ref{sec:litsurveyresults}). Actors may satisfy disclosure or documentation requirements by providing carefully drafted prompts that signal intended values~\cite{bo_disclosures_2024, norval_dicsbydesign_2022}, even when resulting behaviour is difficult to verify or is contingent on other prompting, training, or deployment elements~\cite{de_troya_misabstraction_2025, hopkins2025aisupplychainsemerging, shivagunde2024deconstructingincontextlearningunderstanding}.
Across these risks, the common failure is \textbf{conflating linguistic specification and compliance with behavioural compliance of the system}. If reliance is placed on inspecting and constraining prompt text, rather than evidencing the instructions' behavioural effects, it risks regulating what is legible rather than what is operational; ignoring the fundamental translation challenge between human linguistic interpretation and machine language processing~\cite{li2025humansbrittlelargelanguage, aoyagui_matter_2025, olivier_artificial_2017}.

\subsection{Socio-Technical Implications} \label{sec:discoper}

When system-level instructions are embedded into complex, multi-stakeholder systems~\citep{singh_cobbe_norval_2019}, different implications emerge. These challenges are specifically organisational and operational in nature: aligning prompt-based governance with the priorities of different stakeholders is complicated by different institutional incentives~\cite{kuehnert_who_2025, ali_walking_2023, johnson_legacy_2025}, and by supply chain structures in which responsibilities and capabilities are distributed across multiple actors who often do not share full visibility into each other's constraints~\cite{cobbe_account_2023, widder_dislocatedaccount, hopkins2025aisupplychainsemerging}.

The ease and `low-cost' nature of prompting may cause it to become a \textbf{default intervention} even where alternative mechanisms might prove more robust. If governance is shaped by what is practical to implement within a particular architecture, rather than what is normatively or empirically justified, it may introduce a form of technological determinism~\cite{van2013culture}. 
Yet ignoring the technical aspects in regulation that targets specific technologies is inappropriate. System-level instructions do affect behaviour, albeit in divergent ways~(\S\ref{sec:litsurveyresults}). 
Governance requires operational processes on interpreting prompt disclosures, assessing compliance, detecting and reporting failures, and auditing that links the disclosed artefact to observed behaviour. 

In practice, there are varying approaches as to the 
\textbf{type of system prompt intervention governance}: from high-level guidance on textual content or implementation, to mandates or prohibitions on certain terms, or standardisations. Our findings suggest that prompt-based effects and controls do not transfer reliably across models, training methods, or deployment contexts. Governance mechanisms that rely heavily on text-level prescriptions may produce weak or non-functional safeguards unless coupled with robust evaluation requirements.

Prompt governance must also account for the \textbf{layered instruction stacks that arise across AI supply chains}. Multiple stakeholders can each introduce system-level instructions or adjacent control layers. These layers can interact and conflict depending on technical design choices that are not visible to all actors~\cite{neumann_position_2025, neumann_ai_2026}.
Foundation model layer
constraints are not easy to override, limiting appropriate adaptation. 
This complicates core governance tasks, including accountability attribution~\cite{lima_who_2023, zhang2025getscreditblameattributing, cobbe_account_2023}, disclosure (which layers must be revealed for meaningful oversight)~\cite{norval_disclosure_2022}, and enforcement (which actor has change controls)~\cite{devrio_building_2024, palla_policy-as-prompt_2025}. Treating a single system prompt as the governance object risks misspecifying the system actually being governed.

Governance must address the \textbf{malleability of instructions}. Prompts can be changed (directly or indirectly) across the `prompt stack', be optimised for specific deployers or end-users, and evolve by `system prompt learning'~\cite{Hugging_Face_SystemPromptLearning}. It is important to address this tension, rather than treating static prompts as stable specifications.

\subsection{Societal Implications} \label{sec:discsociety}

These challenges further extend 
to affect how entire communities experience AI systems and the regulatory environments that are meant to govern them.
One reason on why prompt-based governance is attractive is because it appears to \textbf{match the pace of technical change}. System-level instructions can be revised quickly and deployed widely. However, when governance relies on iterative adjustments to prompt text, it risks reproducing a trial-and-error dynamic in governance~\cite{oldenburg2025storiesgovernbyai, neumann2026controls}, with affected communities bearing the costs~\cite{devrio_building_2024, klein_data_2024, Pi_Proctor_2025}. Similar dynamics may also emerge in other textual prompt-based systems beyond GenAI.

Two cross-jurisdictional dynamics are particularly salient. {First}, if prompt requirements are perceived as functional, other jurisdictions may adopt similar approaches to signal regulatory competence or reduce policy development costs. This resembles the so-called ``Brussels Effect''~\cite{bradford2020brussels}, where regulatory requirements become de facto global standards. %
{Second}, because many systems are deployed on global platforms hosting foundation models, \textbf{system prompt constraints can apply across borders}. Where foundational constraints are set to satisfy one jurisdiction, multinational vendors may apply them broadly to reduce operational complexity, meaning that jurisdictions that did not adopt those requirements may nonetheless experience their effects~\cite{info12070275_corpgovernance, davis_eulawspillstous_2024}.
Moreover, there is potential for governmental actors to demand that providers remove, weaken, or add safety restrictions encoded in system-level instructions, using procurement power or regulatory authority as commercial and political leverage over the normative values encoded into deployed models~\cite{newyorkerClaudeAI, openaiAgreementWith, anthropicStatementComments, anthropicWhereThings, gorwa_moderating_2024}.

Affected communities experience these dynamics most directly through shifts in bias and representational harms when prompts encode particular values~\cite{santurkar_whose_2023, diberardino_algorithmic_2024, wang_measuring_2025, 10.1145/3715275.3732015}; trial-and-error deployments when prompt failures are discovered in production~\cite{wiredOpenAIScrambles, thevergeGoogleApologizes, theatlanticElonMusk}; changing system behaviour when prompts are updated without notice~\cite{palla_policy-as-prompt_2025, theguardianGrokundressing, techcrunchOpenAIAdds}, with the possibility of stifling speech~\cite{noels2025largelanguagemodelstalk, buyl_large_2026, marquez_chatgptcensorship} or censorship~\cite{joinrebootAlignmentCensorship, 10.1093/pnasnexus/pgag013}. %
Because prompt patterns can be rapidly copied across systems~\cite{zhang_first_2024}, \textbf{local policy choices can have widespread (global) downstream effects}~\cite{Lima_2023_blaming, agudo_impact_2024}. Regulators should connect prompt artefacts to behavioural evidence that incorporates transparency about prompt changes over time. 

Critically, model capability improvements alone cannot resolve these concerns. The core governance challenge is structural: prompt text is an accessible representation of intended constraints, but not a reliable substitute for evidence about realistic behaviour of layered systems~(\S\ref{sec:litsurveyresults}). If governance instruments treat writing or inspecting a sentence as a primary mechanism of system behavioural control, it risks overstating the impact of prompt-based interventions and understating the need for evaluation and accountability mechanisms~\cite{palla_policy-as-prompt_2025, birhane2024aiauditingbrokenbus, cooper_accountability_2022, poirier_accountable_2022, raza2025responsibledatamodelsusers}.

\subsection{Considerations} \label{sec:considerations} 

The preceding sections document significant challenges with governing AI through natural language instructions. Yet system-level instructions \textit{already} function as control mechanisms across the AI ecosystem--in deployed products, organisational policies, and (emerging) regulatory frameworks. They mediate how millions of people interact with AI systems. Given their documented limitations, a relevant question is not whether to govern instructions, but how to do so meaningfully.

The design and governance of system-level instructions should \textit{evolve beyond textual compliance}. Textual requirements should be paired with targeted evaluation documentation, since blanket requirements across implementations are not suitable. 
Further, these evaluations should be \textit{mindful of the `prompt stack'}, including how instructions by multiple stakeholders interact, and how they are accessed and structured, particularly in related or conflicting prompt configurations.
Towards alleviating ensuing complexity of prompts,
\textit{versioning and documenting} justified prompt modifications %
enables meaningful re-evaluation by multiple stakeholders.
Effective governance further requires cross-disciplinary collaboration, as neither linguistic nor technical expertise alone is sufficient to specify, implement, or assess instruction-based~controls.

For the institutional mechanisms and workflows surrounding prompt development, stakeholders should \textit{resist governance by affordance}: the ease of (re-)writing and accessing a prompt should not determine the scope and intent of governance. Actual behavioural outcomes of the socio-technical system assessed through evaluation regimes must take precedence over easily produced and changed textual artefacts. %
It is imperative to \textit{minimise trial-and-error} deployment. 
As iteratively refining prompts seems like an `easy fix', the pressure to use it as a first-line defence is high; so systematic prompt tests with clear thresholds should be set and met prior to public release. A \textit{standardisation} of any kind should not target words or specific prompt templates unless robustness across implementations is shown. For this, we point to \textit{specialised intermediary roles} 
responsible for translating governance objectives into prompt specifications and validating their behavioural effects~\cite{palla_policy-as-prompt_2025}. %
Lastly, \textit{transparency requirements} %
could specify what information enables meaningful oversight~\cite{cobbe_reviewable_2021}.

\section{Limitations and Future Work} \label{sec:limsfuturework}

Our findings should be interpreted in light of the following limitations. Our systematic review provides structured insight into specific academic claims rather than an exhaustive census of all relevant work. Despite broad searching, additional strands of research may fall outside our corpus. While we did not analyse claim development over time, we still identified considerable categories and convergence.
Longitudinal analysis could trace how evidence and claims shift or converge over time. 

While our analysis did not aim to be exhaustive, we clearly map and synthesis a fragmented landscape that reveals significant limitations in current governance approaches to system-prompt regulation.
To build on our findings, future work could undertake empirical analysis of contested claims in specific governance-relevant settings and examine real-world operational practice to characterise gaps between policy and deployed realities.

\section{Conclusion} \label{sec:conclusion}

System-level instructions have become a governance intervention for GenAI systems because they are legible, editable, and appear to offer direct leverage over model behaviour. 
Yet, our systematic review shows a fragmented research landscape advancing divergent and sometimes contradictory claims about what system-level instructions can achieve, under what conditions, and with what reliability. Our policy analysis similarly indicates that emerging governance instruments treat prompts as artefacts that can support transparency, alignment, and risk mitigation, relying on assumptions of stability and control that are not consistently supported by the evidence.
Our findings demonstrate that governing GenAI by natural language should be approached cautiously: \textbf{\textit{Writing rules that govern machines requires different approaches than writing rules that govern humans.}}

\section*{Acknowledgements}
We thank Bruno Demattos Nogueira, Tamara Stojanovska, and Jasmin Wyss for their valuable insights and assistance in processing our literature review.

\section*{Generative AI Usage Statement} \label{sec:genaistatement}

Asta (allen.ai) was used in the ideation phase of the review to identify links to potentially relevant academic sources, mostly to inform search string development through analysis of frequently used terminology.
Claude Sonnet 4.5 was used as a coding assistant for developing and iterating on Python scripts that (i) extracted codes from PDF and CSV files and (ii) programmatically generated the reference overview table.
The authors also used Claude Sonnet 4.5 for light editorial support and \LaTeX{} formatting advice.

\bibliographystyle{ACM-Reference-Format}
\bibliography{bibs/govern_lang, bibs/govern_final}

\clearpage
\appendix
\onecolumn

\section{Literature Review} \label{app:methodlit}

\paragraph{Final search string.}
Below is the final search string for the literature review. Note that the asterisk (*) represents truncation to capture variant word forms. The proximity operator \texttt{W/n} specifies that terms must appear within \textit{n} words of each other. The search was performed on December 10th, 2025.

{%
\noindent
\fcolorbox{black}{gray!10}{%
\parbox{0.97\textwidth}{%
\footnotesize
\textcolor{keywordcolor}{\texttt{TITLE-ABS}} ( \\
(\textcolor{termcolor}{\texttt{llm*}} \textcolor{operatorcolor}{\texttt{OR}} \textcolor{termcolor}{\texttt{"language model*"}} \textcolor{operatorcolor}{\texttt{OR}} \textcolor{termcolor}{\texttt{gpt*}} \textcolor{operatorcolor}{\texttt{OR}} \textcolor{termcolor}{\texttt{"generative ai"}} \textcolor{operatorcolor}{\texttt{OR}} \textcolor{termcolor}{\texttt{ai}})\\
\quad\textcolor{operatorcolor}{\texttt{AND}} \\ 
\quad\qquad\textcolor{extracolor}{(} \\
\quad\qquad(\textcolor{termcolor}{\texttt{"system prompt*"}} \textcolor{operatorcolor}{\texttt{OR}} \textcolor{termcolor}{\texttt{"system message*"}} \textcolor{operatorcolor}{\texttt{OR}} \textcolor{termcolor}{\texttt{"system instruction*"}})\\
\qquad\textcolor{operatorcolor}{\texttt{OR}} \\ \textcolor{extracolor}{(} \\
\qquad((\textcolor{termcolor}{\texttt{instruction}} \textcolor{operatorcolor}{\texttt{OR}} \textcolor{termcolor}{\texttt{rule}} \textcolor{operatorcolor}{\texttt{OR}} \textcolor{termcolor}{\texttt{prompt*}}) \textcolor{keywordcolor}{\texttt{W/0}} (\textcolor{termcolor}{\texttt{following}} \textcolor{operatorcolor}{\texttt{OR}} \textcolor{termcolor}{\texttt{adherence}} \textcolor{operatorcolor}{\texttt{OR}} \textcolor{termcolor}{\texttt{compliance}}))
\textcolor{keywordcolor}{\texttt{W/5}} (\textcolor{termcolor}{\texttt{evaluat*}} \textcolor{operatorcolor}{\texttt{OR}} \textcolor{termcolor}{\texttt{benchmark*}} \textcolor{operatorcolor}{\texttt{OR}} \textcolor{termcolor}{\texttt{assess*}} \textcolor{operatorcolor}{\texttt{OR}} \textcolor{termcolor}{\texttt{measur*}})\\
\textcolor{extracolor}{)}\\
\qquad\textcolor{operatorcolor}{\texttt{OR}} ((\textcolor{termcolor}{\texttt{prompt*}}) \textcolor{keywordcolor}{\texttt{W/5}} (\textcolor{termcolor}{\texttt{policy}} \textcolor{operatorcolor}{\texttt{OR}} \textcolor{termcolor}{\texttt{governance}}))\\
\quad\textcolor{extracolor}{)}\\)
\\\textcolor{operatorcolor}{\texttt{AND}} \textcolor{keywordcolor}{\texttt{PUBYEAR}} > 2021
\\\textcolor{operatorcolor}{\texttt{AND}} \textcolor{keywordcolor}{\texttt{PUBYEAR}} < 2026
\\\textcolor{operatorcolor}{\texttt{AND}} \textcolor{keywordcolor}{\texttt{LIMIT-TO}}(\textcolor{termcolor}{\texttt{LANGUAGE}}, \textcolor{termcolor}{\texttt{"English"}})
}}
\captionof{figure}{Search string for literature review.}
\label{box:searchstring}
}%

\paragraph{Annotator collaboration.}
Two annotators conducted the full-text review of 373 papers. The two annotators independently piloted the extraction procedure on ten papers each, aligned on definitions and recoding conventions, and then split the remaining corpus. Disagreements and edge cases were resolved through discussion, with consensus meetings used to clarify inclusion and coding decisions. As a final consistency check, all excluded full-text papers were assessed or re-assessed by the primary annotator. 

\section{Policy Documents Review} \label{app:methodlegal}

We document the process used to identify policy and governance documents that explicitly address system-level instructions. Unlike the academic literature review (\S\ref{sec:litsurveymethods}), this was not designed as a systematic review. Policy artefacts are dispersed across jurisdictions and institutions, are published in heterogeneous formats, and often change or move across websites. We therefore conducted a scoping-style, practice-oriented search aimed at locating a best-efforts snapshot of publicly accessible governance documents that engage system-level instructions as governance objects.

\paragraph{Search strategy and sources.}
We used an iterative query strategy that combined three term families: model terms (e.g., \texttt{llm*}, \texttt{foundation model*}, \texttt{general-purpose ai}), system-level instruction terms (e.g., \texttt{system prompt*}, \texttt{system instruction*}, \texttt{developer message*}, \texttt{model specification}), and governance terms (e.g., \texttt{govern*}, \texttt{regulat*}, \texttt{procure*}, \texttt{risk}, \texttt{assess*}, \texttt{audit*}). 
We executed these queries across (i) official government websites (e.g., US Federal Register, GOV.UK), (ii) national and supranational policy portals and repositories (e.g., Australian Policy Online, IAPP, OECD.AI, Overton) (iii) general web search using domain- and filetype-restricted queries (e.g., \texttt{site:*.gov filetype:pdf}). We complemented these searches with manual snowballing by iteratively following relevant hyperlinks and internal citations from candidate documents to locate additional materials. %

\paragraph{Screening and selection.} Given the breadth of this landscape, we screened approximately 300 documents at title and executive-summary level for potential relevance. We retained 54 documents for closer review because they contained at least one plausible reference to system-level instructions or closely related constructs and made governance-relevant claims (e.g., transparency expectations, documentation duties, evaluation requirements, security controls). We then read these 54 documents in full and applied narrower inclusion criteria, resulting in a final set of eight documents that substantively address system-level instructions.

\paragraph{Inclusion and exclusion criteria.} We included documents that (i) contain an explicit reference to what we define as system-level instructions (including terms such as system prompts, developer messages, model specifications, or equivalent descriptions), and (ii) articulate governance-facing expectations, requirements, or guidance related to system-level instructions  (e.g., transparency, documentation, evaluation, security controls). We excluded documents that addressed only end-user prompting practices, as well as academic or policy research outputs that did not themselves function as guidance or requirements.

\clearpage
\section{Coding Analysis} \label{app:coding}

As we engaged in an open-ended coding process, we could find multiple claims per paper, and multiple codes per claim. Therefore our categories are inherently diverse, showing a more fine-grained landscape of papers, claims, and goals. Partly due to this open process, some of our codes could have belonged in multiple categories as they discussed or mentioned multiple goals. To determine a primary category for each code, we noted down possible category matches for each and went through the claims again to determine their overarching primary goals, discussed its inclusion, and agreed on a primary category. We noted down a justification for each claim for why it was included in its primary category (see \autoref{tab:claim_tag_comprehensive}).

\footnotesize
\begin{longtable}{p{0.13\textwidth} p{0.10\textwidth} p{0.45\textwidth} r r r}
    \caption{Comprehensive analysis of claim tags with categories, justifications, and statistics}\label{tab:claim_tag_comprehensive} \\
    \toprule
    \textbf{Tag} & \textbf{Other Categories} & \textbf{Justification} & \textbf{Count} & \textbf{Pos} & \textbf{Neg} \\
    \midrule
    \endfirsthead

    \multicolumn{6}{c}{{\tablename\ \thetable{} -- continued from previous page}} \\
    \toprule
    \textbf{Tag} & \textbf{Other Categories} & \textbf{Justification} & \textbf{Count} & \textbf{Supp} & \textbf{Crit} \\
    \midrule
    \endhead

    \midrule
    \multicolumn{6}{r}{{Continued on next page}} \\
    \endfoot

    \bottomrule
    \endlastfoot
    \midrule
    \multicolumn{6}{l}{\textbf{Accessibility}} \\
    \midrule
    Ease of Use via Natural Language & - & Claims address how easily system prompts can be implemented or modified in systems, making model control more accessible to diverse stakeholders & 46 & 39 & 7 \\
    Transparency of the System & -  & Claims examine whether system prompts make the system itself more transparent or interpretable by enabling stakeholders to understand intended behaviour & 17 & 10 & 7 \\
    Explainability Enhancement & - & Claims focus specifically on whether system prompts make the system more explainable to users or auditors & 3 & 3 & 0 \\
    Understandable Optimal Instructions & - & Claims address the accessibility implications when prompts are not human-readable or interpretable & 3 & 0 & 3 \\
    \midrule
    \multicolumn{6}{l}{\textbf{Adaptability}} \\
    \midrule
    Adaptation to Domain/Context & Performance & Claims address adapting models to specific domains or contexts through system prompts, not evaluating performance specifically within those domains & 178 & 172 & 6 \\
    Behaviour Modification & Performance & Claims examine inducing specific behaviour through system prompts, distinct from evaluating performance on those behaviour & 159 & 148 & 11 \\
    Policy/Guideline Implementation & Alignment & Claims focus on adapting models to guidelines set by developing stakeholders rather than aligning to broader values, though these are closely related & 112 & 96 & 16 \\
    Persona/Role Implementation & -  & Claims examine adapting systems to fulfill specific personas or roles through system prompts & 83 & 67 & 16 \\
    Agent Creation & - & Claims address transforming LLMs into agents specifically through system prompt configuration & 44 & 41 & 3 \\
    System Transformation & -  & Claims examine transforming systems to perform different functions, like LLM as a Judge or classifiers, through system prompts & 38 & 36 & 2 \\
    \midrule
    \multicolumn{6}{l}{\textbf{Alignment}} \\
    \midrule
    Alignment & - & Claims address whether system-level instructions align model behaviour with stakeholder values and goals & 48 & 29 & 19 \\
    Responsible Guardrails & Safety & Claims primarily examine value-based constraints (e.g., child safety, content policies) rather than technical security mechanisms & 44 & 36 & 8 \\
    User Safety & Security & Claims focus on aligning models with safety values for end-users rather than implementing security mechanisms & 38 & 31 & 7 \\
    Hierarchy Establishment & Performance & Claims address whether instruction hierarchies successfully establish control structures that prioritise developer intentions, treating this as alignment to deployer/developer goals & 26 & 15 & 11 \\
    Explicit Value Specification & - & Claims examine setting and prioritizing explicit values (e.g., honesty over helpfulness) through system-level instructions & 20 & 17 & 3 \\
    Bias Reduction & Performance & Claims focus on bias reduction as an alignment objective rather than as performance optimization on benchmarks & 19 & 10 & 9 \\
    HHH Principles & - & Claims evaluate effects on the helpfulness, harmlessness, and honesty principles & 11 & 8 & 3 \\
    behaviour Alignment with Human Expectations & Adaptability & Claims examine whether user or researcher expectations when writing instructions align with actual model behaviour, rather than just whether behaviour changes occur & 11 & 1 & 10 \\
    Diversity Enhancement & Performance & Claims primarily treat diversity of outputs as an alignment goal rather than a performance optimization target & 7 & 2 & 5 \\
    \midrule
    \multicolumn{6}{l}{\textbf{Auditability}} \\
    \midrule
    Effect Transferability & - & Claims examine whether auditing goals can even be achieved, since transferable effects enable single evaluations to inform broader assessments while non-transferable effects require different auditing approaches & 104 & 39 & 65 \\
    Oversight Requirement & -  & Claims address oversight of the prompt creation and modification process, not oversight of the system itself & 28 & 25 & 3 \\
    Evaluation Necessity & -  & Claims focus on evaluating prompt effects and comparing different prompts, not evaluating the system broadly & 23 & 21 & 2 \\
    Transparency of the Prompt & -  & Claims address transparency about how prompts are made and changed over time, and who can see them when & 10 & 6 & 4 \\
    \midrule
    \multicolumn{6}{l}{\textbf{Implementation}} \\
    \midrule
    Iterative Improvement & - & Claims address meta-level approaches for how to make prompts better through iterative processes & 74 & 72 & 2 \\
    Machine-Engineered Solutions & -  & Claims address meta-level approaches for how prompts can be engineered through human expertise & 64 & 61 & 3 \\
    Human-Engineered Solutions & -  & Claims examine meta-level approaches for how prompts can be made better through automated methods & 48 & 43 & 5 \\
    Static Layer Consistency & - & Claims examine the goal of applying instructions at the right time or in the right way. This concerns prompt implementation timing rather than system stability, as it addresses stability of the prompt itself not the system & 37 & 11 & 26 \\
    Are secure & Security & Claims assess the security of prompts themselves in current deployment practices, not the security of the system & 13 & 1 & 12 \\
    Over-reliance Risk & Performance & Claims address approaches to using prompts and their functionality rather than the system functionality they influence & 12 & 1 & 11 \\
    Lengthy Instructions & -  & Claims identify characteristics of the current prompt implementation landscape through documentation of typical lengths & 9 & 7 & 2 \\
    Similarity in Instructions & - & Claims identify characteristics of the current prompt implementation landscape through documentation of commonalities across deployed prompts & 2 & 2 & 0 \\
    \midrule
    \multicolumn{6}{l}{\textbf{Performance}} \\
    \midrule
    Performance Boosting & - & Claims directly evaluate performance improvements or degradations & 172 & 150 & 22 \\
    Improvement through Long-Form/Specificity & -  & Claims examine performance of different lengths of prompts rather than implementation requirements for achieving good performance, making this a system goal not a prompt goal & 57 & 25 & 32 \\
    Cost Efficiency & -  & Claims focus on cheaper system operation compared to alternatives, not on making the prompt itself better or more efficient & 39 & 28 & 11 \\
    Accuracy Improvement & - & Claims specifically measure how accuracy is impacted, making accuracy improvement the inherent goal & 36 & 31 & 5 \\
    Superior to Existing Methods & - & Claims compare methods for achieving system goals, focusing on choosing optimal approaches rather than improving prompts themselves & 32 & 23 & 9 \\
    Incorporation in Other Methods & - & Claims examine what performance contributions system prompts make when integrated into other methodological frameworks & 14 & 14 & 0 \\
    Interaction Quality Enhancement & Alignment & Claims evaluate performance through user interaction quality metrics rather than focusing on aligning with user values specifically & 9 & 7 & 2 \\
    Influence from Other Methods & Implementation & Claims examine how system performance is affected by other methods (training, architectures), emphasizing performance outcomes rather than prompt implementation techniques & 9 & 9 & 0 \\
    \midrule
    \multicolumn{6}{l}{\textbf{Security}} \\
    \midrule
    Resistant to Jailbreaking & -  & Claims address the vulnerability that system prompts introduce by being jailbreakable, or claims that they are no longer, focusing on how systems are compromised through system prompts & 69 & 7 & 62 \\
    Safeguard-Ability & -  & Claims focus on using system prompts to keep systems safer from security threats & 56 & 47 & 9 \\
    Safety (System) & -  & Claims address the security properties of the system itself as a goal & 40 & 32 & 8 \\
    Confidential Data Protection & Implementation & Claims examine the security dangers inherent in system prompts containing intellectual property and the need to defend against extraction & 27 & 11 & 16 \\
    Require Safeguarding & -  & Claims identify system prompts as vulnerabilities requiring safeguarding to enhance overall system security & 23 & 0 & 23 \\
    Vulnerable to Jailbreaking & -  & Claims examine system prompts as tools for jailbreaking systems, ultimately the goal is making systems more secure from prompt injection attacks through the analysis & 16 & 0 & 16 \\
    Detectable & Implementation & Claims examine how system prompts make systems or models identifiable, treating this primarily as a security issue rather than an implementation characteristic & 11 & 3 & 8 \\
    Prompt Injection Protection & -  & Claims focus on using system prompts to make the system more secure, not on securing the prompts themselves & 6 & 2 & 4 \\
    Performance-Safety Tradeoff & Performance & Claims primarily evaluate safety tradeoffs, prioritizing security assessment over performance optimization & 5 & 3 & 2 \\
    \midrule
    \multicolumn{6}{l}{\textbf{Stability}} \\
    \midrule
    Hierarchical Instruction Following & - & Claims examine maintaining consistent, predictable behaviour according to designed priority structures, regardless of varying or conflicting inputs & 87 & 45 & 42 \\
    Effect Stability & - & Claims directly ask whether prompts produce stable effects in the system across contexts and time & 84 & 12 & 72 \\
    Uniform Language Adherence & Performance, Adaptability & Claims examine differences in how systems react to various phrasings or language, showing how stability varies across prompt language & 53 & 1 & 52 \\
    Consistency Improvement & -  & Claims focus on the consistency of outputs, directly serving the stability goal & 26 & 21 & 5 \\
    Robustness Enhancement & Security, Performance & Claims examine what makes outputs more robust, prioritizing this stability dimension over performance metrics or security properties & 21 & 20 & 1 \\
\end{longtable}

\clearpage
\section{Example Claims} \label{app:exclaims}

\begin{table*}[h]
    \footnotesize
    \caption{Exemplary supportive and critical perspectives on system-level instruction goals.}
    \label{tab:claim_examples}
    \resizebox{\textwidth}{!}{
    \begin{tabularx}{\textwidth}{>{\hsize=0.3\hsize}X >{\hsize=0.25\hsize}X >{\hsize=2.45\hsize}X} 
    \toprule
    \textbf{Goal} & \textbf{Stance} & \textbf{Exemplary quotes} \\
    \midrule
    \multirow{2}{=}{\textit{Alignment}}
    & \textbf{Supportive}
    & \textit{``We have put careful effort into creating system prompts that give instructions to prevent harmful, offensive, or disrespectful responses to respect people’s values and provide reliable results''}~\citep{yusuf_rag-based_2024}.
    \vspace{1mm}\\
    & \textbf{Critical}
    & \textit{``Our study shows that strategically prompting LLMs to create malicious LLM-based CAIs to extract personal information is not only highly effective but also alarmingly simple. Our findings confirm that crafting a single malicious system prompt is sufficient to transform an LLM-based CAI into an effective tool for personal data extraction''}~\citep{zhan_malicious_2025}\\
    \midrule
    \multirow{2}{=}{\textit{Accessibility}}
    & \textbf{Supportive} 
    & \textit{``Users can reshape the assistant's behavior simply by editing the system prompt---adjusting its tone, logic, or domain specificity using natural language. This makes the assistant highly flexible and easy torepurpose''}~\citep{le_conversational_2025}
    \vspace{1mm}\\
    & \textbf{Critical}
    & \textit{``Users cannot reliably anticipate emergent behaviors from system prompts alone. The opaqueness of this mechanism means users are essentially operating blind during the design process, discovering problems only after deployment through trial and error''}~\citep{karny_neural_2025}\\
    \midrule
    \multirow{2}{=}{\textit{Adaptability}}
    & \textbf{Supportive} 
    & \textit{``We find that system prompts that strongly discourage or encourage the behavior are able to define the behavior in our setting almost entirely, leading to nearly (but not exactly) 0\% or 100\% rates of misaligned behavior''}~\citep{scheurer_large_2024}
    \vspace{1mm}\\
    & \textbf{Critical}
    & \textit{``Using system prompts to shape persona is particularly brittle, where adversarial instructions frequently 'break character' and lead to generic 'helpful assistant'-style behavior''}~\citep{maiya_open_2025}\\
    \midrule
    \multirow{2}{=}{\textit{Performance}}
    & \textbf{Supportive} 
    & \textit{``Optimizing the system prompt to include task details and an advisory about structured input further elevates GPT-4.1's accuracy by an additional 10-13 percentage points''} \citep{braun_hidden_2025}
    \vspace{1mm}\\
    & \textbf{Critical} 
    & \textit{``We observed that more complex and lengthy system prompts lead to degraded generation quality''} \citep{shayegani_measurement_2025}\\
    \midrule
    \multirow{2}{=}{\textit{Stability}}
    & \textbf{Supportive} 
    & \textit{``Complementors guide the AI model's reasoning in a way that is optimized to make AI model outputs more consistent in terms of content, format, and syntax''} \citep{heimburg_complementor_2025}
    \vspace{1mm}\\
    & \textbf{Critical} 
    & \textit{``When the chatbot drifts away from its system prompts that stipulate safety aspects, it becomes more susceptible to jailbreaking and more prone to hallucinations''} \citep{li_measuring_2024}\\
    \midrule
    \multirow{2}{=}{\textit{Security}}
    & \textbf{Supportive} 
    & \textit{``[P]otential system prompt leakage risks, model developers have implemented strategies to prevent leakage, primarily by disabling LLMs from repeating their context when encountering known attack patterns''} \citep{cao_you_2025}
    \vspace{1mm}\\
    & \textbf{Critical} 
    & \textit{``System prompts can be revealed by an adversary despite existing defenses in place''} \citep{zhang_effective_2024}\\
    \midrule
    \midrule
    \multirow{2}{=}{\textit{Implementation}}
    & \textbf{Supportive} 
    & \textit{``An expertly written system prompt [...] could be critical in getting quality responses from large back end LLMs''} \citep{levin_has_2025}
    \vspace{1mm}\\
    & \textbf{Critical} 
    & \textit{``[E]xcessive reliance on this functionality raises the risk of manipulation by malicious actors who can exploit it''} \citep{chan_detection_2024}\\
    \midrule
    \multirow{2}{=}{\textit{Auditability}}
    & \textbf{Supportive} 
    & \textit{``Based on these findings, we urge developers and prompt engineers to rigorously evaluate the safety and robustness of prompt-based systems, incorporating adversarial testing as a standard part of agent design and deployment''} \citep{kang_doppelganger_2025}
    \vspace{1mm}\\
    & \textbf{Critical} 
    & \textit{``If every minor change requires extensive documentation, it adds unnecessary overhead. Yet, if changes are not sufficiently documented, it becomes increasingly difficult to monitor and understand the impact of these modifications''} \citep{palla_policy-as-prompt_2025}\\
    \bottomrule
    \end{tabularx}
    }
\end{table*}

\clearpage
\section{References for Typology} \label{app:corpusrefs}

\footnotesize

\begin{longtable}{p{0.1\textwidth} p{0.1\textwidth} p{0.70\textwidth}}
    \caption{References for all categories by stance.}\label{tab:group_citations} \\
    \toprule
    \textbf{Category} & \textbf{Stance} & \textbf{References} \\
    \midrule
    \endfirsthead

    \multicolumn{3}{c}{{\tablename\ \thetable{} -- continued from previous page}} \\
    \toprule
    \textbf{Category} & \textbf{Stance} & \textbf{References} \\
    \midrule
    \endhead

    \midrule
    \multicolumn{3}{r}{{Continued on next page}} \\
    \endfoot

    \bottomrule
    \endlastfoot
    \multirow{2}{=}{\textit{Alignment}}
    & \textbf{Supportive}
    & \citep{NEURIPS2024_de7b9910}, \citep{alan_improving_2025}, \citep{alhujaili_fine-tuning_2025}, \citep{alkhowaiter_mind_2025}, \citep{ben-zion_detecting_2025}, \citep{bo_disclosures_2024}, \citep{boie_evaluating_2025}, \citep{cai_simulation_2025}, \citep{cohen_information_2025}, \citep{debenedetti_dataset_2024}, \citep{doudkin_spark_2025}, \citep{ermolaeva_how_2024}, \citep{furmakiewicz_design_2024}, \citep{gallego_configurable_2024}, \citep{gallego_metasc_2025}, \citep{greenblatt_alignment_2024}, \citep{guo_review_2024}, \citep{halil_llm_2025}, \citep{hazan_security_2025}, \citep{heimburg_complementor_2025}, \citep{hu_droj_2024}, \citep{huang_how_2024}, \citep{iyenghar_empirical_2025}, \citep{jeong_system_2025}, \citep{kahan_artificial_2025}, \citep{katz_segment-based_2024}, \citep{kholkar_policy-as-prompt_2025}, \citep{khoo_minorbench_2025}, \citep{kim_debiasing_2025}, \citep{krishna_d-rex_2025}, \citep{kucharavy_adapting_2024}, \citep{kumar_no_2025}, \citep{lee_aligning_2024}, \citep{lee_data-model_2025}, \citep{lei_offtopiceval_2025}, \citep{li_securitylingua_2025}, \citep{li_test-time_2025}, \citep{liang_integrating_2025}, \citep{liang_kda_2025}, \citep{lin_unlocking_2023}, \citep{liu_generative_2025}, \citep{lu_rolemrc_2025}, \citep{mukherjee_orca_2023}, \citep{mushtaq_towards_2025}, \citep{mushtaq_worldview-bench_2025}, \citep{neelou_a2as_2025}, \citep{neumann_position_2025}, \citep{ogundoyin_unsafe_2025}, \citep{palla_policy-as-prompt_2025}, \citep{pauli_measuring_2025}, \citep{peng_navigating_2024}, \citep{qiu_training_2024}, \citep{qu_beyond_2025}, \citep{samaei_epidemiqs_2025}, \citep{scheurer_large_2024}, \citep{scholl_students_2025}, \citep{sharma_spml_2024}, \citep{sharma_sysformer_2025}, \citep{shayegani_measurement_2025}, \citep{shi_navigating_2024}, \citep{sievers_using_2025}, \citep{skripko_instruction-following_2025}, \citep{tan_inoculation_2025}, \citep{tuggener_role-playing_2024}, \citep{vachharajani_transformer_2025}, \citep{vidgen_simplesafetytests_2024}, \citep{wallace_instruction_2024}, \citep{wang_assessing_2025}, \citep{wang_attngcg_2024}, \citep{wang_illusion_2025}, \citep{wang_ip_2025}, \citep{wei_evaluating_2024}, \citep{wu_instructional_2025}, \citep{xie_defending_2023}, \citep{xu_mixture--instructions_2025}, \citep{xu_tool_2023}, \citep{yan_codeif_2025}, \citep{yang_promptcos_2025}, \citep{yu_high_2024}, \citep{yu_rewardanything_2025}, \citep{yusuf_rag-based_2024}, \citep{zeng_who_2025}, \citep{zhan_malicious_2025}, \citep{zhang_controllable_2025}, \citep{zhang_iheval_2025}, \citep{zhang_iopo_2025}, \citep{zhang_personaagent_2025}, \citep{zhao_adversarial_2024}, \citep{zheng_reasoning_2025} \\
    & \textbf{Critical}
    & \citep{adiletta_rubber_2025}, \citep{almasi_alignment_2025}, \citep{arike_technical_2025}, \citep{ben-zion_detecting_2025}, \citep{cai_simulation_2025}, \citep{chiang_why_2025}, \citep{chiu_dailydilemmas_2025}, \citep{dodgson_establishing_2024}, \citep{greenblatt_alignment_2024}, \citep{han_beyond_2024}, \citep{huang_how_2024}, \citep{jia_one_2025}, \citep{karny_neural_2025}, \citep{khoo_minorbench_2025}, \citep{kim_debiasing_2025}, \citep{kirichenko_abstentionbench_2025}, \citep{kumar_can_2025}, \citep{lei_offtopiceval_2025}, \citep{leidinger_how_2024}, \citep{li_evaluating_2023}, \citep{li_gspr_2025}, \citep{li_system_2025}, \citep{liang_integrating_2025}, \citep{liang_kda_2025}, \citep{lu_rolemrc_2025}, \citep{modi_assessing_2025}, \citep{mu_can_2024}, \citep{mushtaq_towards_2025}, \citep{mushtaq_worldview-bench_2025}, \citep{naik_agentmisalignment_2025}, \citep{navneet_rethinking_2025}, \citep{neumann_position_2025}, \citep{palla_policy-as-prompt_2025}, \citep{peng_navigating_2024}, \citep{pham_cain_2025}, \citep{qin_sysbench_2024}, \citep{qu_beyond_2025}, \citep{russell_assessing_2025}, \citep{scheurer_large_2024}, \citep{scholl_students_2025}, \citep{shi_navigating_2024}, \citep{sovrano_can_2025}, \citep{tuggener_role-playing_2024}, \citep{wallace_instruction_2024}, \citep{zeng_who_2025}, \citep{zhan_malicious_2025}, \citep{zhang_effective_2024}, \citep{zhang_iheval_2025}, \citep{zheng_when_2024} \\
    \midrule
    \multirow{2}{=}{\textit{Security}}
    & \textbf{Supportive}
    & \citep{NEURIPS2024_de7b9910}, \citep{agarwal_prompt_2024}, \citep{ben-zion_detecting_2025}, \citep{chan_detection_2024}, \citep{chen_flexllm_2024}, \citep{chen_pitfalls_2025}, \citep{chua_flexible_2025}, \citep{das_system_2025}, \citep{debenedetti_dataset_2024}, \citep{gakh_enhancing_2025}, \citep{gallego_metasc_2025}, \citep{gosmar_prompt_2025}, \citep{greenblatt_alignment_2024}, \citep{guo_review_2024}, \citep{heimburg_complementor_2025}, \citep{huang_dynamic_2025}, \citep{hui_span_2024}, \citep{kang_doppelganger_2025}, \citep{kumar_no_2025}, \citep{lei_offtopiceval_2025}, \citep{li_securitylingua_2025}, \citep{liu_secinfer_2025}, \citep{maiya_open_2025}, \citep{neelou_a2as_2025}, \citep{ogundoyin_unsafe_2025}, \citep{pape_prompt_2025}, \citep{peng_navigating_2024}, \citep{peng_repeatleakage_2025}, \citep{pfister_gandalf_2025}, \citep{sharma_spml_2024}, \citep{siska_attentiondefense_2025}, \citep{thakur_defending_2025}, \citep{wallace_instruction_2024}, \citep{wang_attngcg_2024}, \citep{wang_ip_2025}, \citep{wang_protect_2025}, \citep{wu_instructional_2025}, \citep{wu_slip_2025}, \citep{xie_defending_2023}, \citep{yang_promptcos_2025}, \citep{zeng_who_2025}, \citep{zhang_controllable_2025}, \citep{zhang_effective_2024}, \citep{zhang_first_2024}, \citep{zhao_adversarial_2024}, \citep{zheng_improved_2024}, \citep{zheng_reasoning_2025}, \citep{zhu_autodan_2023}, \citep{zhuang_proxyprompt_2025}, \citep{zou_is_2024} \\
    & \textbf{Critical}
    & \citep{adiletta_rubber_2025}, \citep{agarwal_prompt_2024}, \citep{bernstein_trust_2025}, \citep{cao_you_2025}, \citep{chao_jailbreakbench_2024}, \citep{chen_multi-turn_2024}, \citep{chen_pitfalls_2025}, \citep{chiang_large_2024}, \citep{chiang_why_2025}, \citep{daniel_impact_2024}, \citep{das_system_2025}, \citep{gakh_enhancing_2025}, \citep{gallotta_prompt_2025}, \citep{greenblatt_alignment_2024}, \citep{hagendorff_large_2025}, \citep{he_membership_2025}, \citep{hui_span_2024}, \citep{jiang_promptkeeper_2025}, \citep{kang_doppelganger_2025}, \citep{krishna_d-rex_2025}, \citep{kucharavy_adapting_2024}, \citep{li_evaluating_2023}, \citep{li_measuring_2024}, \citep{li_rethinking_2025}, \citep{li_system_2025}, \citep{liang_kda_2025}, \citep{liang_universal_2024}, \citep{maloyan_adversarial_2025}, \citep{nazir_better_2025}, \citep{nie_leakagent_2025}, \citep{ogundoyin_unsafe_2025}, \citep{pan_latentqa_2024}, \citep{pape_prompt_2025}, \citep{pasquini_llmmap_2025}, \citep{peng_repeatleakage_2025}, \citep{pfister_gandalf_2025}, \citep{pham_cain_2025}, \citep{reddy_autoadv_2025}, \citep{salem_maatphor_2023}, \citep{sam_predicting_2025}, \citep{sharma_sysformer_2025}, \citep{siska_attentiondefense_2025}, \citep{song_early_2025}, \citep{thakur_defending_2025}, \citep{vidgen_simplesafetytests_2024}, \citep{wallace_instruction_2024}, \citep{wang_attngcg_2024}, \citep{wang_closer_2025}, \citep{wang_ip_2025}, \citep{wang_protect_2025}, \citep{wu_instructional_2025}, \citep{wu_slip_2025}, \citep{xie_defending_2023}, \citep{xie_logic-rl_2025}, \citep{yang_promptcos_2025}, \citep{yu_assessing_2024}, \citep{zhang_agent_2025}, \citep{zhang_effective_2024}, \citep{zhang_enja_2024}, \citep{zhang_extracting_2024}, \citep{zhang_first_2024}, \citep{zheng_improved_2024}, \citep{zhu_autodan_2023}, \citep{zou_is_2024} \\
    \midrule
    \multirow{2}{=}{\textit{Performance}}
    & \textbf{Supportive}
    & \citep{adebimpe_sbash_2025}, \citep{ahlgren_assisting_2025}, \citep{ali_developing_2025}, \citep{amjad_agentic_2025}, \citep{arai_enhancing_2025}, \citep{arif_agentic_2025}, \citep{battle_unreasonable_2024}, \citep{ben-zion_detecting_2025}, \citep{bernstein_trust_2025}, \citep{bethany_camouflage_2025}, \citep{bo_disclosures_2024}, \citep{braun_hidden_2025}, \citep{buyl_building_2025}, \citep{cai_simulation_2025}, \citep{cao_you_2025}, \citep{challagundla_si-agent_2025}, \citep{chantangphol_finmind-y-me_2025}, \citep{chen_parapo_2025}, \citep{chen_pitfalls_2025}, \citep{cheng_err_2025}, \citep{choi_system_2025}, \citep{chon_multifaceted_2025}, \citep{cohen_information_2025}, \citep{collini_c2hlsc_2025}, \citep{costa_instruction-level_2025}, \citep{dan_educhat_2023}, \citep{diao_guidebench_2025}, \citep{dodgson_establishing_2024}, \citep{dong_toward_2024}, \citep{doudkin_spark_2025}, \citep{fallah_automating_2024}, \citep{farid_hf_detox_2025}, \citep{fuchs_using_2024}, \citep{furmakiewicz_design_2024}, \citep{gallego_configurable_2024}, \citep{gallego_metasc_2025}, \citep{gao_meta_2024}, \citep{gosmar_prompt_2025}, \citep{halil_llm_2025}, \citep{han_exploring_2025}, \citep{harb_towards_2025}, \citep{hofmann_beyond_2025}, \citep{hu_droj_2024}, \citep{huang_agent_2024}, \citep{huang_how_2024}, \citep{hui_span_2024}, \citep{iadisernia_prompting_2025}, \citep{iyenghar_empirical_2025}, \citep{jeong_system_2025}, \citep{jia_one_2025}, \citep{jiang_promptkeeper_2025}, \citep{kahan_artificial_2025}, \citep{katz_segment-based_2024}, \citep{kholkar_policy-as-prompt_2025}, \citep{khoo_minorbench_2025}, \citep{kucharavy_adapting_2024}, \citep{kumar_no_2025}, \citep{lan_llm-based_2024}, \citep{lashuel_llm_2024}, \citep{lee_aligning_2024}, \citep{lee_data-model_2025}, \citep{lee_large_2025}, \citep{lei_offtopiceval_2025}, \citep{levin_has_2025}, \citep{li_graphif_2025}, \citep{li_llm-match_2025}, \citep{li_matryoshka_2025}, \citep{li_securitylingua_2025}, \citep{li_test-time_2025}, \citep{li_when_2025}, \citep{liang_integrating_2025}, \citep{liang_kda_2025}, \citep{liu_analyzing_2025}, \citep{liu_generative_2025}, \citep{liu_notebook_nodate}, \citep{liu_oraplan-sql_2025}, \citep{liu_secinfer_2025}, \citep{lu_rolemrc_2025}, \citep{malik_scaffolding_2025}, \citep{maloyan_adversarial_2025}, \citep{mcginness_large_2025}, \citep{mu_policy_2025}, \citep{muti_customizing_2024}, \citep{neelou_a2as_2025}, \citep{nguyen_software_2025}, \citep{ocker_tulip_2024}, \citep{palla_policy-as-prompt_2025}, \citep{panda_causal_2025}, \citep{pape_prompt_2025}, \citep{peng_navigating_2024}, \citep{piterbarg_diff_2024}, \citep{polignano_unraveling_2024}, \citep{qiu_training_2024}, \citep{ramos_transparent_2024}, \citep{reddy_autoadv_2025}, \citep{rumiantsau_beyond_2024}, \citep{samaei_epidemiqs_2025}, \citep{sami_adaptive_2025}, \citep{scholl_students_2025}, \citep{sharma_sysformer_2025}, \citep{shen_optimizing_2025}, \citep{siska_attentiondefense_2025}, \citep{sovrano_can_2025}, \citep{tan_inoculation_2025}, \citep{thelwall_evaluating_2024}, \citep{uittenhove_metacognitive_2025}, \citep{vachharajani_pjmathematician_2025}, \citep{vachharajani_transformer_2025}, \citep{verdaguer-gonzalez_boosting_2025}, \citep{vijayvargiya_efficient_2025}, \citep{wang_assessing_2025}, \citep{wang_codeif-bench_2025}, \citep{wang_meeseeks_2025}, \citep{wang_protect_2025}, \citep{wang_rnr_2024}, \citep{wang_towards_2025}, \citep{wang_we_2025}, \citep{watts_framework_2025}, \citep{wei_evaluating_2024}, \citep{wongso_genup_2025}, \citep{wroblewska_integrating_2025}, \citep{wu_automedprompt_2025}, \citep{wu_comparative_2024}, \citep{wu_instructional_2025}, \citep{xie_defending_2023}, \citep{xie_logic-rl_2025}, \citep{xu_mixture--instructions_2025}, \citep{xu_procut_2025}, \citep{xu_tool_2023}, \citep{yan_refutebench_2024}, \citep{yang_promptcos_2025}, \citep{yao_leveraging_2024}, \citep{yao_retroformer_2024}, \citep{yu_high_2024}, \citep{yu_rewardanything_2025}, \citep{yuan_dmt-rolebench_2025}, \citep{yusuf_rag-based_2024}, \citep{zeng_evaluating_2024}, \citep{zhang_agentic_2025}, \citep{zhang_automated_2025}, \citep{zhang_autonomous_2023}, \citep{zhang_leveraging_2025}, \citep{zhang_personaagent_2025}, \citep{zhang_sprig_2024}, \citep{zhao_adversarial_2024}, \citep{zhao_large_2026}, \citep{zhao_saber_2025}, \citep{zheng_improved_2024}, \citep{zheng_reasoning_2025}, \citep{zheng_when_2024}, \citep{zhuang_proxyprompt_2025}, \citep{zou_bisup_2024}, \citep{zou_is_2024} \\
    & \textbf{Critical}
    & \citep{cao_you_2025}, \citep{challagundla_si-agent_2025}, \citep{chen_flexllm_2024}, \citep{chen_pitfalls_2025}, \citep{chiang_large_2024}, \citep{chiang_why_2025}, \citep{das_system_2025}, \citep{duan_hierarchical_2025}, \citep{fallah_automating_2024}, \citep{fuchs_using_2024}, \citep{greenblatt_alignment_2024}, \citep{han_exploring_2025}, \citep{hofmann_beyond_2025}, \citep{hu_droj_2024}, \citep{iadisernia_prompting_2025}, \citep{jaroslawicz_how_2025}, \citep{khoo_minorbench_2025}, \citep{kirichenko_abstentionbench_2025}, \citep{kumar_no_2025}, \citep{lee_large_2025}, \citep{li_effects_2024}, \citep{li_measuring_2024}, \citep{li_system_2025}, \citep{li_when_2025}, \citep{liang_integrating_2025}, \citep{liang_kda_2025}, \citep{lichkovski_eu-agent-bench_2025}, \citep{liu_analyzing_2025}, \citep{maiya_open_2025}, \citep{marquardt_fine-tuning_2025}, \citep{mu_closer_2025}, \citep{mukherjee_orca_2023}, \citep{neumann_position_2025}, \citep{ng_prism_2025}, \citep{nguyen_software_2025}, \citep{ocker_tulip_2024}, \citep{palla_policy-as-prompt_2025}, \citep{peng_navigating_2024}, \citep{pfister_gandalf_2025}, \citep{piterbarg_diff_2024}, \citep{russell_assessing_2025}, \citep{schuurmans_autoregressive_2024}, \citep{shayegani_measurement_2025}, \citep{shi_llm-augmented_2025}, \citep{sun_beyond_2024}, \citep{thelwall_evaluating_2025}, \citep{tuggener_role-playing_2024}, \citep{vijayvargiya_efficient_2025}, \citep{wang_codeif-bench_2025}, \citep{wang_illusion_2025}, \citep{wang_meeseeks_2025}, \citep{wang_towards_2025}, \citep{wei_evaluating_2024}, \citep{wongso_genup_2025}, \citep{wroblewska_integrating_2025}, \citep{xu_mixture--instructions_2025}, \citep{xu_procut_2025}, \citep{zhang_enja_2024}, \citep{zhang_iheval_2025}, \citep{zhang_sprig_2024}, \citep{zheng_when_2024}, \citep{zhu_relayattention_2024}, \citep{zou_bisup_2024} \\
    \midrule
    \multirow{2}{=}{\textit{Adaptability}}
    & \textbf{Supportive}
    & \citep{NEURIPS2024_de7b9910}, \citep{adebimpe_sbash_2025}, \citep{ahlgren_assisting_2025}, \citep{akaybicen_machine_2024}, \citep{alan_improving_2025}, \citep{alhujaili_fine-tuning_2025}, \citep{ali_developing_2025}, \citep{alkhowaiter_mind_2025}, \citep{almasi_alignment_2025}, \citep{amjad_agentic_2025}, \citep{arai_enhancing_2025}, \citep{arif_agentic_2025}, \citep{arike_technical_2025}, \citep{ayyamperumal_current_2024}, \citep{becker_boardwalk_2025}, \citep{ben-zion_detecting_2025}, \citep{bethany_camouflage_2025}, \citep{bo_disclosures_2024}, \citep{boie_evaluating_2025}, \citep{braun_hidden_2025}, \citep{buyl_building_2025}, \citep{cai_simulation_2025}, \citep{cao_you_2025}, \citep{challagundla_si-agent_2025}, \citep{chan_detection_2024}, \citep{chantangphol_finmind-y-me_2025}, \citep{chen_diahalu_2024}, \citep{chen_parapo_2025}, \citep{chen_pitfalls_2025}, \citep{cheng_err_2025}, \citep{chiang_large_2024}, \citep{choi_system_2025}, \citep{chon_multifaceted_2025}, \citep{cohen_information_2025}, \citep{collini_c2hlsc_2025}, \citep{costa_instruction-level_2025}, \citep{dan_educhat_2023}, \citep{debenedetti_dataset_2024}, \citep{diao_guidebench_2025}, \citep{dodgson_establishing_2024}, \citep{doudkin_spark_2025}, \citep{fallah_automating_2024}, \citep{farid_hf_detox_2025}, \citep{fuchs_using_2024}, \citep{furmakiewicz_design_2024}, \citep{gakh_enhancing_2025}, \citep{gallego_configurable_2024}, \citep{gallego_metasc_2025}, \citep{gao_meta_2024}, \citep{gosmar_prompt_2025}, \citep{greenblatt_alignment_2024}, \citep{hagendorff_large_2025}, \citep{halil_llm_2025}, \citep{han_exploring_2025}, \citep{harada_when_2025}, \citep{harb_towards_2025}, \citep{hazan_security_2025}, \citep{heimburg_complementor_2025}, \citep{hofmann_beyond_2025}, \citep{huang_agent_2024}, \citep{huang_how_2024}, \citep{hui_span_2024}, \citep{iyenghar_empirical_2025}, \citep{jeong_system_2025}, \citep{jia_one_2025}, \citep{kahan_artificial_2025}, \citep{karny_neural_2025}, \citep{katz_segment-based_2024}, \citep{kempny_qualipilot_2025}, \citep{kholkar_policy-as-prompt_2025}, \citep{khoo_minorbench_2025}, \citep{kim_debiasing_2025}, \citep{kirichenko_abstentionbench_2025}, \citep{kucharavy_adapting_2024}, \citep{kumar_can_2025}, \citep{kumar_no_2025}, \citep{lan_llm-based_2024}, \citep{lashuel_llm_2024}, \citep{le_conversational_2025}, \citep{lee_aligning_2024}, \citep{lee_data-model_2025}, \citep{lee_large_2025}, \citep{lei_offtopiceval_2025}, \citep{levin_has_2025}, \citep{li_effects_2024}, \citep{li_graphif_2025}, \citep{li_llm-match_2025}, \citep{li_matryoshka_2025}, \citep{li_measuring_2024}, \citep{li_rethinking_2025}, \citep{li_securitylingua_2025}, \citep{li_system_2025}, \citep{li_test-time_2025}, \citep{li_when_2025}, \citep{liang_integrating_2025}, \citep{lichkovski_eu-agent-bench_2025}, \citep{lin_unlocking_2023}, \citep{liu_analyzing_2025}, \citep{liu_generative_2025}, \citep{liu_notebook_nodate}, \citep{liu_observability_2024}, \citep{liu_oraplan-sql_2025}, \citep{liu_secinfer_2025}, \citep{lu_rolemrc_2025}, \citep{macnaught_designing_2025}, \citep{malik_scaffolding_2025}, \citep{mcginness_large_2025}, \citep{meglio_large_2025}, \citep{modi_assessing_2025}, \citep{mu_policy_2025}, \citep{mushtaq_towards_2025}, \citep{muti_customizing_2024}, \citep{naik_agentmisalignment_2025}, \citep{neelou_a2as_2025}, \citep{neumann_position_2025}, \citep{nguyen_software_2025}, \citep{ocker_tulip_2024}, \citep{ogundoyin_unsafe_2025}, \citep{palla_policy-as-prompt_2025}, \citep{panda_causal_2025}, \citep{pape_prompt_2025}, \citep{park_truebench_2025}, \citep{pasquini_llmmap_2025}, \citep{pauli_measuring_2025}, \citep{peng_repeatleakage_2025}, \citep{polignano_unraveling_2024}, \citep{qin_sysbench_2024}, \citep{ramos_transparent_2024}, \citep{rumiantsau_beyond_2024}, \citep{russell_assessing_2025}, \citep{salem_maatphor_2023}, \citep{samaei_epidemiqs_2025}, \citep{sami_adaptive_2025}, \citep{scheurer_large_2024}, \citep{scholl_students_2025}, \citep{schuurmans_autoregressive_2024}, \citep{sharma_spml_2024}, \citep{sharma_sysformer_2025}, \citep{shayegani_measurement_2025}, \citep{shen_optimizing_2025}, \citep{sievers_using_2025}, \citep{skripko_instruction-following_2025}, \citep{sovrano_can_2025}, \citep{sun_beyond_2024}, \citep{thelwall_evaluating_2024}, \citep{tuggener_role-playing_2024}, \citep{uittenhove_metacognitive_2025}, \citep{vachharajani_pjmathematician_2025}, \citep{vachharajani_transformer_2025}, \citep{verdaguer-gonzalez_boosting_2025}, \citep{vijayvargiya_efficient_2025}, \citep{wallace_instruction_2024}, \citep{wand_three-phases-lora_2024}, \citep{wang_codeif-bench_2025}, \citep{wang_illusion_2025}, \citep{wang_ip_2025}, \citep{wang_meeseeks_2025}, \citep{wang_protect_2025}, \citep{wang_rnr_2024}, \citep{wang_towards_2025}, \citep{wang_we_2025}, \citep{wei_evaluating_2024}, \citep{wongso_genup_2025}, \citep{wu_automedprompt_2025}, \citep{wu_comparative_2024}, \citep{wu_slip_2025}, \citep{xie_logic-rl_2025}, \citep{xu_mixture--instructions_2025}, \citep{xu_procut_2025}, \citep{xu_tool_2023}, \citep{yang_promptcos_2025}, \citep{yao_leveraging_2024}, \citep{yao_retroformer_2024}, \citep{yu_high_2024}, \citep{yu_rewardanything_2025}, \citep{yuan_dmt-rolebench_2025}, \citep{yusuf_rag-based_2024}, \citep{zeng_evaluating_2024}, \citep{zhan_malicious_2025}, \citep{zhang_agentic_2025}, \citep{zhang_automated_2025}, \citep{zhang_decoding_2025}, \citep{zhang_iheval_2025}, \citep{zhang_leveraging_2025}, \citep{zhang_mapro_2025}, \citep{zhang_personaagent_2025}, \citep{zhang_sprig_2024}, \citep{zhao_adversarial_2024}, \citep{zhao_large_2026}, \citep{zhao_saber_2025}, \citep{zheng_reasoning_2025}, \citep{zheng_when_2024}, \citep{zhu_autodan_2023} \\
    & \textbf{Critical}
    & \citep{adiletta_rubber_2025}, \citep{almasi_alignment_2025}, \citep{chen_parapo_2025}, \citep{chiang_large_2024}, \citep{das_system_2025}, \citep{gallotta_prompt_2025}, \citep{hagendorff_large_2025}, \citep{han_beyond_2024}, \citep{iadisernia_prompting_2025}, \citep{kang_doppelganger_2025}, \citep{katz_segment-based_2024}, \citep{kucharavy_adapting_2024}, \citep{kumar_can_2025}, \citep{lashuel_llm_2024}, \citep{lei_offtopiceval_2025}, \citep{li_effects_2024}, \citep{li_gspr_2025}, \citep{li_measuring_2024}, \citep{li_rethinking_2025}, \citep{liang_integrating_2025}, \citep{lichkovski_eu-agent-bench_2025}, \citep{maiya_open_2025}, \citep{naik_agentmisalignment_2025}, \citep{palla_policy-as-prompt_2025}, \citep{scheurer_large_2024}, \citep{shi_llm-augmented_2025}, \citep{sun_beyond_2024}, \citep{wang_illusion_2025}, \citep{wang_ip_2025}, \citep{wang_rnr_2024}, \citep{weesep_exploring_2025}, \citep{xie_logic-rl_2025}, \citep{zeng_who_2025}, \citep{zhang_agent_2025}, \citep{zhang_enja_2024}, \citep{zheng_when_2024} \\
    \midrule
    \multirow{2}{=}{\textit{Accessibility}}
    & \textbf{Supportive}
    & \citep{ahlgren_assisting_2025}, \citep{bo_disclosures_2024}, \citep{cao_you_2025}, \citep{challagundla_si-agent_2025}, \citep{chen_flexllm_2024}, \citep{choi_system_2025}, \citep{cohen_information_2025}, \citep{costa_instruction-level_2025}, \citep{gallego_configurable_2024}, \citep{gallego_metasc_2025}, \citep{gosmar_prompt_2025}, \citep{harb_towards_2025}, \citep{hazan_security_2025}, \citep{heimburg_complementor_2025}, \citep{kholkar_policy-as-prompt_2025}, \citep{le_conversational_2025}, \citep{lee_data-model_2025}, \citep{levin_has_2025}, \citep{li_test-time_2025}, \citep{modi_assessing_2025}, \citep{mu_policy_2025}, \citep{navneet_rethinking_2025}, \citep{palla_policy-as-prompt_2025}, \citep{panda_causal_2025}, \citep{pape_prompt_2025}, \citep{pham_cain_2025}, \citep{qu_beyond_2025}, \citep{ramos_transparent_2024}, \citep{shen_optimizing_2025}, \citep{sun_beyond_2024}, \citep{verdaguer-gonzalez_boosting_2025}, \citep{wallace_instruction_2024}, \citep{wongso_genup_2025}, \citep{xu_tool_2023}, \citep{yu_rewardanything_2025}, \citep{zeng_evaluating_2024}, \citep{zhan_malicious_2025}, \citep{zhang_mapro_2025}, \citep{zhao_large_2026}, \citep{zheng_reasoning_2025}, \citep{zhuang_proxyprompt_2025} \\
    & \textbf{Critical}
    & \citep{battle_unreasonable_2024}, \citep{cao_you_2025}, \citep{challagundla_si-agent_2025}, \citep{chan_detection_2024}, \citep{gallotta_prompt_2025}, \citep{karny_neural_2025}, \citep{li_rethinking_2025}, \citep{liu_observability_2024}, \citep{neumann_position_2025}, \citep{palla_policy-as-prompt_2025}, \citep{wallace_instruction_2024}, \citep{wu_slip_2025}, \citep{zhang_personaagent_2025} \\
    \midrule
    \multirow{2}{=}{\textit{Stability}}
    & \textbf{Supportive}
    & \citep{adebimpe_sbash_2025}, \citep{alkhowaiter_mind_2025}, \citep{arike_technical_2025}, \citep{cao_you_2025}, \citep{chen_pitfalls_2025}, \citep{choi_system_2025}, \citep{cobbina_where_2025}, \citep{cohen_information_2025}, \citep{collini_c2hlsc_2025}, \citep{costa_instruction-level_2025}, \citep{debenedetti_dataset_2024}, \citep{dong_toward_2024}, \citep{fallah_automating_2024}, \citep{gallego_configurable_2024}, \citep{gosmar_prompt_2025}, \citep{greenblatt_alignment_2024}, \citep{harada_when_2025}, \citep{heimburg_complementor_2025}, \citep{jeong_system_2025}, \citep{jia_one_2025}, \citep{katz_segment-based_2024}, \citep{kim_debiasing_2025}, \citep{kumar_no_2025}, \citep{li_graphif_2025}, \citep{li_measuring_2024}, \citep{li_system_2025}, \citep{liu_generative_2025}, \citep{liu_notebook_nodate}, \citep{lu_rolemrc_2025}, \citep{mehralian_codealignbench_2025}, \citep{mu_closer_2025}, \citep{neumann_position_2025}, \citep{palla_policy-as-prompt_2025}, \citep{polignano_unraveling_2024}, \citep{qi_constraint_2025}, \citep{qiu_training_2024}, \citep{samaei_epidemiqs_2025}, \citep{sami_adaptive_2025}, \citep{scholl_students_2025}, \citep{schuurmans_autoregressive_2024}, \citep{sharma_spml_2024}, \citep{sharma_sysformer_2025}, \citep{skripko_instruction-following_2025}, \citep{sovrano_can_2025}, \citep{thakur_defending_2025}, \citep{tuggener_role-playing_2024}, \citep{verdaguer-gonzalez_boosting_2025}, \citep{wallace_instruction_2024}, \citep{wang_codeif-bench_2025}, \citep{wang_illusion_2025}, \citep{wang_ip_2025}, \citep{wang_rnr_2024}, \citep{wu_instructional_2025}, \citep{xu_mixture--instructions_2025}, \citep{xu_procut_2025}, \citep{xu_tool_2023}, \citep{yan_codeif_2025}, \citep{yan_refutebench_2024}, \citep{yusuf_rag-based_2024}, \citep{zeng_who_2025}, \citep{zhang_automated_2025}, \citep{zhang_controllable_2025}, \citep{zhang_iheval_2025}, \citep{zhang_iopo_2025}, \citep{zheng_reasoning_2025}, \citep{zou_is_2024} \\
    & \textbf{Critical}
    & \citep{almasi_alignment_2025}, \citep{arike_technical_2025}, \citep{becker_boardwalk_2025}, \citep{bernstein_trust_2025}, \citep{cai_simulation_2025}, \citep{cao_you_2025}, \citep{chen_flexllm_2024}, \citep{chen_pitfalls_2025}, \citep{chiang_large_2024}, \citep{chiang_why_2025}, \citep{chiu_dailydilemmas_2025}, \citep{chon_multifaceted_2025}, \citep{cobbina_where_2025}, \citep{daniel_impact_2024}, \citep{das_system_2025}, \citep{debenedetti_dataset_2024}, \citep{diao_guidebench_2025}, \citep{diederich_rule_2025}, \citep{doudkin_spark_2025}, \citep{duan_hierarchical_2025}, \citep{farid_hf_detox_2025}, \citep{furmakiewicz_design_2024}, \citep{gao_meta_2024}, \citep{gosmar_prompt_2025}, \citep{greenblatt_alignment_2024}, \citep{hackmann_word_2024}, \citep{han_beyond_2024}, \citep{harada_when_2025}, \citep{hazan_security_2025}, \citep{iadisernia_prompting_2025}, \citep{iyenghar_empirical_2025}, \citep{jaroslawicz_how_2025}, \citep{kang_doppelganger_2025}, \citep{karny_neural_2025}, \citep{khoo_minorbench_2025}, \citep{kucharavy_adapting_2024}, \citep{kumar_can_2025}, \citep{lee_aligning_2024}, \citep{lei_offtopiceval_2025}, \citep{leidinger_how_2024}, \citep{levin_has_2025}, \citep{li_gspr_2025}, \citep{li_measuring_2024}, \citep{li_rethinking_2025}, \citep{li_when_2025}, \citep{liang_integrating_2025}, \citep{liu_analyzing_2025}, \citep{liu_generative_2025}, \citep{lu_rolemrc_2025}, \citep{marquardt_fine-tuning_2025}, \citep{mehralian_codealignbench_2025}, \citep{mu_can_2024}, \citep{muti_customizing_2024}, \citep{navneet_rethinking_2025}, \citep{neumann_position_2025}, \citep{nguyen_software_2025}, \citep{palla_policy-as-prompt_2025}, \citep{park_truebench_2025}, \citep{pauli_measuring_2025}, \citep{qi_constraint_2025}, \citep{qin_infobench_2024}, \citep{qin_sysbench_2024}, \citep{qiu_training_2024}, \citep{sam_predicting_2025}, \citep{schlatter_shutdown_2025}, \citep{scholl_students_2025}, \citep{shi_navigating_2024}, \citep{siska_attentiondefense_2025}, \citep{sun_beyond_2024}, \citep{thakur_defending_2025}, \citep{thelwall_evaluating_2025}, \citep{uittenhove_metacognitive_2025}, \citep{verdaguer-gonzalez_boosting_2025}, \citep{wand_three-phases-lora_2024}, \citep{wang_attngcg_2024}, \citep{wang_illusion_2025}, \citep{wang_meeseeks_2025}, \citep{wang_we_2025}, \citep{weesep_exploring_2025}, \citep{wu_instructional_2025}, \citep{xie_logic-rl_2025}, \citep{xu_mixture--instructions_2025}, \citep{yan_codeif_2025}, \citep{yan_refutebench_2024}, \citep{zeng_who_2025}, \citep{zhan_malicious_2025}, \citep{zhang_automated_2025}, \citep{zhang_decoding_2025}, \citep{zhang_effective_2024}, \citep{zhang_first_2024}, \citep{zhang_iheval_2025}, \citep{zheng_when_2024}, \citep{zou_is_2024} \\
    \midrule
    \multirow{2}{=}{\textit{Implementation}}
    & \textbf{Supportive}
    & \citep{adebimpe_sbash_2025}, \citep{ahlgren_assisting_2025}, \citep{ali_developing_2025}, \citep{amjad_agentic_2025}, \citep{arai_enhancing_2025}, \citep{arif_agentic_2025}, \citep{ayyamperumal_current_2024}, \citep{battle_unreasonable_2024}, \citep{bethany_camouflage_2025}, \citep{braun_hidden_2025}, \citep{cai_simulation_2025}, \citep{challagundla_si-agent_2025}, \citep{chen_flexllm_2024}, \citep{cheng_err_2025}, \citep{choi_system_2025}, \citep{cohen_information_2025}, \citep{costa_instruction-level_2025}, \citep{diao_guidebench_2025}, \citep{diederich_rule_2025}, \citep{doudkin_spark_2025}, \citep{farid_hf_detox_2025}, \citep{furmakiewicz_design_2024}, \citep{gallego_metasc_2025}, \citep{greenblatt_alignment_2024}, \citep{hagendorff_large_2025}, \citep{hazan_security_2025}, \citep{heimburg_complementor_2025}, \citep{huang_dynamic_2025}, \citep{huang_how_2024}, \citep{jones_teaching_2023}, \citep{kahan_artificial_2025}, \citep{karny_neural_2025}, \citep{katz_segment-based_2024}, \citep{kim_debiasing_2025}, \citep{kumar_no_2025}, \citep{le_conversational_2025}, \citep{lee_aligning_2024}, \citep{lee_data-model_2025}, \citep{lee_large_2025}, \citep{lei_offtopiceval_2025}, \citep{levin_has_2025}, \citep{li_graphif_2025}, \citep{li_matryoshka_2025}, \citep{li_test-time_2025}, \citep{liang_integrating_2025}, \citep{liang_kda_2025}, \citep{liu_generative_2025}, \citep{liu_notebook_nodate}, \citep{liu_observability_2024}, \citep{liu_oraplan-sql_2025}, \citep{liu_secinfer_2025}, \citep{lu_rolemrc_2025}, \citep{macnaught_designing_2025}, \citep{meglio_large_2025}, \citep{mu_policy_2025}, \citep{nguyen_software_2025}, \citep{palla_policy-as-prompt_2025}, \citep{pape_prompt_2025}, \citep{park_truebench_2025}, \citep{peng_navigating_2024}, \citep{piterbarg_diff_2024}, \citep{qiu_training_2024}, \citep{ramos_transparent_2024}, \citep{reddy_autoadv_2025}, \citep{salem_maatphor_2023}, \citep{sami_adaptive_2025}, \citep{schuurmans_autoregressive_2024}, \citep{sharma_spml_2024}, \citep{sharma_sysformer_2025}, \citep{shen_optimizing_2025}, \citep{sievers_using_2025}, \citep{siska_attentiondefense_2025}, \citep{thelwall_evaluating_2025}, \citep{tuggener_role-playing_2024}, \citep{uittenhove_metacognitive_2025}, \citep{vachharajani_pjmathematician_2025}, \citep{vachharajani_transformer_2025}, \citep{verdaguer-gonzalez_boosting_2025}, \citep{wang_protect_2025}, \citep{wang_we_2025}, \citep{wongso_genup_2025}, \citep{wu_automedprompt_2025}, \citep{wu_slip_2025}, \citep{xie_defending_2023}, \citep{xu_procut_2025}, \citep{xu_tool_2023}, \citep{yang_promptcos_2025}, \citep{yao_leveraging_2024}, \citep{yao_retroformer_2024}, \citep{ye_chunkattention_2024}, \citep{yu_high_2024}, \citep{yusuf_rag-based_2024}, \citep{zeng_evaluating_2024}, \citep{zhan_malicious_2025}, \citep{zhang_agentic_2025}, \citep{zhang_automated_2025}, \citep{zhang_first_2024}, \citep{zhang_mapro_2025}, \citep{zhang_personaagent_2025}, \citep{zhang_sprig_2024}, \citep{zhao_large_2026}, \citep{zhu_relayattention_2024}, \citep{zhuang_proxyprompt_2025}, \citep{zou_is_2024} \\
    & \textbf{Critical}
    & \citep{almasi_alignment_2025}, \citep{arai_enhancing_2025}, \citep{battle_unreasonable_2024}, \citep{cai_simulation_2025}, \citep{challagundla_si-agent_2025}, \citep{chan_detection_2024}, \citep{costa_instruction-level_2025}, \citep{das_system_2025}, \citep{diederich_rule_2025}, \citep{gallego_metasc_2025}, \citep{he_membership_2025}, \citep{jia_one_2025}, \citep{kirichenko_abstentionbench_2025}, \citep{kucharavy_adapting_2024}, \citep{li_gspr_2025}, \citep{li_measuring_2024}, \citep{li_securitylingua_2025}, \citep{li_system_2025}, \citep{liang_kda_2025}, \citep{malik_scaffolding_2025}, \citep{nazir_better_2025}, \citep{neelou_a2as_2025}, \citep{neumann_position_2025}, \citep{palla_policy-as-prompt_2025}, \citep{peng_repeatleakage_2025}, \citep{piterbarg_diff_2024}, \citep{qiu_training_2024}, \citep{salem_maatphor_2023}, \citep{sami_adaptive_2025}, \citep{scheurer_large_2024}, \citep{schuurmans_autoregressive_2024}, \citep{sharma_sysformer_2025}, \citep{tuggener_role-playing_2024}, \citep{wang_protect_2025}, \citep{wang_towards_2025}, \citep{wongso_genup_2025}, \citep{yan_refutebench_2024}, \citep{zhang_first_2024}, \citep{zhang_personaagent_2025}, \citep{zheng_when_2024}, \citep{zou_bisup_2024} \\
    \midrule
    \multirow{2}{=}{\textit{Auditability}}
    & \textbf{Supportive}
    & \citep{agarwal_prompt_2024}, \citep{arike_technical_2025}, \citep{battle_unreasonable_2024}, \citep{chen_parapo_2025}, \citep{cheng_err_2025}, \citep{chiang_large_2024}, \citep{choi_system_2025}, \citep{cobbina_where_2025}, \citep{costa_instruction-level_2025}, \citep{diederich_rule_2025}, \citep{doudkin_spark_2025}, \citep{gallego_metasc_2025}, \citep{gosmar_prompt_2025}, \citep{greenblatt_alignment_2024}, \citep{heimburg_complementor_2025}, \citep{huang_agent_2024}, \citep{jeong_system_2025}, \citep{jones_teaching_2023}, \citep{kang_doppelganger_2025}, \citep{kempny_qualipilot_2025}, \citep{kholkar_policy-as-prompt_2025}, \citep{kumar_can_2025}, \citep{lee_aligning_2024}, \citep{lee_data-model_2025}, \citep{levin_has_2025}, \citep{li_graphif_2025}, \citep{li_system_2025}, \citep{liu_notebook_nodate}, \citep{liu_oraplan-sql_2025}, \citep{lu_rolemrc_2025}, \citep{malik_scaffolding_2025}, \citep{muti_customizing_2024}, \citep{navneet_rethinking_2025}, \citep{neelou_a2as_2025}, \citep{neumann_position_2025}, \citep{ocker_tulip_2024}, \citep{palla_policy-as-prompt_2025}, \citep{pape_prompt_2025}, \citep{peng_navigating_2024}, \citep{piterbarg_diff_2024}, \citep{sam_predicting_2025}, \citep{schuurmans_autoregressive_2024}, \citep{takeshita_gengo_2025}, \citep{wallace_instruction_2024}, \citep{wang_we_2025}, \citep{wongso_genup_2025}, \citep{xie_defending_2023}, \citep{xu_mixture--instructions_2025}, \citep{xu_tool_2023}, \citep{yang_promptcos_2025}, \citep{yu_rewardanything_2025}, \citep{zhang_sprig_2024}, \citep{zhao_adversarial_2024}, \citep{zheng_reasoning_2025}, \citep{zhuang_proxyprompt_2025}, \citep{zou_is_2024} \\
    & \textbf{Critical}
    & \citep{NEURIPS2024_de7b9910}, \citep{ali_developing_2025}, \citep{almasi_alignment_2025}, \citep{amjad_agentic_2025}, \citep{arai_enhancing_2025}, \citep{arike_technical_2025}, \citep{battle_unreasonable_2024}, \citep{braun_hidden_2025}, \citep{buyl_building_2025}, \citep{chan_detection_2024}, \citep{cobbina_where_2025}, \citep{daniel_impact_2024}, \citep{das_system_2025}, \citep{diao_guidebench_2025}, \citep{duan_hierarchical_2025}, \citep{gakh_enhancing_2025}, \citep{gao_meta_2024}, \citep{greenblatt_alignment_2024}, \citep{hazan_security_2025}, \citep{iadisernia_prompting_2025}, \citep{jaroslawicz_how_2025}, \citep{jia_one_2025}, \citep{karny_neural_2025}, \citep{kholkar_policy-as-prompt_2025}, \citep{khoo_minorbench_2025}, \citep{lashuel_llm_2024}, \citep{lei_offtopiceval_2025}, \citep{levin_has_2025}, \citep{li_evaluating_2023}, \citep{li_gspr_2025}, \citep{li_measuring_2024}, \citep{li_system_2025}, \citep{liang_universal_2024}, \citep{lu_rolemrc_2025}, \citep{mcginness_large_2025}, \citep{mehralian_codealignbench_2025}, \citep{mu_closer_2025}, \citep{mukherjee_orca_2023}, \citep{neumann_position_2025}, \citep{palla_policy-as-prompt_2025}, \citep{panda_causal_2025}, \citep{pape_prompt_2025}, \citep{park_truebench_2025}, \citep{peng_repeatleakage_2025}, \citep{qin_infobench_2024}, \citep{qin_sysbench_2024}, \citep{schlatter_shutdown_2025}, \citep{sharma_spml_2024}, \citep{thelwall_evaluating_2025}, \citep{wang_meeseeks_2025}, \citep{wang_towards_2025}, \citep{weesep_exploring_2025}, \citep{xu_tool_2023}, \citep{zeng_evaluating_2024}, \citep{zhang_automated_2025}, \citep{zhang_effective_2024}, \citep{zhang_iheval_2025}, \citep{zheng_when_2024}, \citep{zhu_autodan_2023} \\
\end{longtable}

\end{document}